\documentclass[preprint,aps,amssymb,showpacs,superscriptaddress]{revtex4}
\usepackage{graphicx}
\begin{document}
\preprint{IMSc/2003/04/06}
\preprint{CGPG--03/10--5}
\preprint{AEI--2003--085}

\title{Homogeneous Loop Quantum Cosmology:\\ The Role of the Spin
Connection}

\author{Martin Bojowald}
\email{mabo@aei.mpg.de}
\affiliation{Max-Planck-Institut f\"ur Gravitationsphysik,
Albert-Einstein-Institut,\\
Am M\"uhlenberg 1, D-14476 Golm, Germany}
\affiliation{Center for Gravitational Physics and Geometry,\\
The Pennsylvania State University,\\
104 Davey Lab, University Park, PA 16802, USA}
\author{Ghanashyam Date}
\email{shyam@imsc.res.in}
\affiliation{The Institute of Mathematical Sciences\\
CIT Campus, Chennai-600 113, INDIA.}
\author{Kevin Vandersloot}
\email{kfvander@gravity.psu.edu}
\affiliation{Center for Gravitational Physics and Geometry,\\
The Pennsylvania State University,\\
104 Davey Lab, University Park, PA 16802, USA}

\newtheorem{Def}{Definition}

\newcommand{\Case}[2]{{\textstyle \frac{#1}{#2}}}
\newcommand{\lP}{\ell_{\mathrm P}}
\newcommand{\HE}{H^{(\mathrm E)}}
\newcommand{\hatHE}{\hat{H}^{(\mathrm E)}}

\newcommand{\md}{{\mathrm{d}}}
\newcommand{\tr}{\mathop{\mathrm{tr}}}
\newcommand{\sgn}{\mathop{\mathrm{sgn}}}

\begin{abstract}
Homogeneous cosmological models with non-vanishing intrinsic
curvature require a special treatment when they are quantized with
loop quantum cosmological methods. Guidance from the full theory
which is lost in this context can be replaced by two criteria for an
acceptable quantization, admissibility of a continuum approximation
and local stability. A quantization of the corresponding Hamiltonian
constraints is presented and shown to lead to a locally stable,
non-singular evolution compatible with almost classical behavior at
large volume. As an application, the Bianchi IX model and its
modified behavior close to its classical singularity is explored.
\end{abstract}

\pacs{0460P, 0460K, 9880H}

\maketitle

\section{Introduction}

Since observations show that space is homogeneous at large scales to a
very good approximation, homogeneous models have been studied for many
decades in classical and quantum cosmology \cite{Misner}. They provide
insights into the behavior of our universe while avoiding all the
complicated field theoretic details of full gravity by their
restriction to finitely many degrees of freedom. Despite the huge
reduction by infinitely many degrees of freedom, a large variety of
different models is left which allow a detailed investigation of
various non-trivial issues such as: Dirac observables, the embedding of
symmetric models (e.g., isotropic) into less symmetric ones, dynamics,
the approach to classical singularities and cosmological
phenomenology.

Loop quantum cosmological methods for those models have recently been
developed \cite{cosmoI,cosmoII,cosmoIII,cosmoIV} and shown to simplify
considerably after a diagonalization of the connection and triad
degrees of freedom \cite{HomCosmo}. As an example the dynamics of the
Bianchi I model has been studied and shown to be singularity-free. In
this case the Hamiltonian constraint resembles that of the full theory
\cite{QSDI} (albeit it is much simpler), and thus the quantization can
be regarded as a reliable test. It turned out that the extension from
the isotropic case \cite{IsoCosmo,Sing} to this anisotropic case,
requires certain features which are in fact present in quantum
geometry. In particular, the mechanism of a singularity-free
evolution in anisotropic models relies on the fact that the evolution
extends through the classical singularity which can be identified with
a submanifold in minisuperspace. This requires the classical
singularity to lie in the interior of minisuperspace, rather than at
the boundary (including infinity). Here it is essential that quantum
geometry is based on densitized triad variables where in fact the
Kasner singularity is in the interior (all densitized triad components
being zero), while it would be on the infinite boundary in co-triad
variables. The isotropic case is not sensitive to this issue since
both densitized and co-triad variables have the classical singularity
in the interior of minisuperspace (here, it is essential to use any
(co-)triad rather than metric variables); thus the existence of the
extension of the methods to a homogeneous model is non-trivial. An
extension to the full theory of the general class of inhomogeneous
models seems much more complicated at this point.  However, it has
been conjectured \cite{BKL} that at the classical level, the most
general homogeneous behavior is that of the Bianchi-IX model and
furthermore that it describes the approach to the singularities even
of {\em inhomogeneous} models. It is therefore important to study the
more general Bianchi class A models.

Homogeneous models other than Bianchi I, however, present an
additional complication since they have non-zero intrinsic curvature
and, as a consequence, their spin connection cannot be zero. This is
in contrast even to the full theory where the spin connection can be
made arbitrarily small locally by choosing appropriate coordinates. In
a homogeneous model the freedom of choosing coordinates is restricted
to those which preserve (manifest) homogeneity, and this implies that
the spin connection is a covariant object and that it has to have a
certain size in a given model. This has to be taken into account
properly when one quantizes the Hamiltonian constraint. In particular,
in a classical regime only the extrinsic curvature can be assumed to
be small, but not necessarily the intrinsic curvature which determines
the spin connection. This is essential to understand the semiclassical
behavior.

Since a special treatment is required which is not necessary in the
full theory, one has to be more careful when interpreting the
results. It introduces more quantization ambiguities which have to be
shown not to influence main results. On the other hand, more
possibilities for phenomenology emerge which, at least qualitatively,
can often be seen to be insensitive to ambiguities.

Having a more distant relation to the full theory, it is helpful
to have a set of admissibility criteria. One such criterion is of
course to accommodate a semiclassical approximation to the quantum
dynamics. Since a semiclassical description is based on continuum
geometry while the quantum dynamics is in terms of discrete quantum
geometry, this criterion is formulated here in terms of a {\em
continuum approximation}. A further criterion is provided by the
requirement of a {\em locally stable evolution} \cite{Stab}.  Such a
requirement arises because the evolution equation derived along the
lines of \cite{cosmoIII,cosmoIV} is usually a high order difference
equation which at large volume is well approximated by the second
order Wheeler--DeWitt differential equation. Hence, one can always
construct solutions by choosing initial data, at large volume, to be
that provided by solutions of the Wheeler--DeWitt
equation. Perturbation of such an initial data will generate solutions
which will also include surplus solutions of the high order difference
equation. Generically these will have Planck scale oscillations. There
is then the possibility that such extra solutions can become dominant
under the evolution. In fact these are expected to become dominant as
one gets to smaller volume since there are huge differences between
the continuum and the discrete formulations in the Planck
regime. However as one evolves to larger volumes, the perturbations
must not grow too rapidly since this would imply domination by
solutions with Planck scale oscillation even in the classical
regime. The requirement we are looking for, local stability,
prohibits this behavior by demanding that the local behavior of
solutions to the difference equation around a large value of the
evolution parameter is not exponentially increasing.  In this paper we
present a quantization of the Hamiltonian constraint for Bianchi
models with non-zero intrinsic curvature which fulfills this condition
and has the correct semiclassical limit. The first quantization given
in \cite{cosmoIII} was not admissible in this sense; thus, the
quantization given there is valid only for the Bianchi I model (for
which we have the same quantization here).

It turns out that the local stability condition is selective: it
requires all roots of a high order polynomial to have unit norm which
is not easy to achieve randomly. The selectivity is increased by the
fact that the same strategy of quantizing a Hamiltonian constraint
should work in all homogeneous models, using different procedures in
different models would imply that not all of them can be related to
the full theory.

The stability condition can be side-stepped by quantizing connection
components in the constraint by hand such that only a second order
difference equation results. While this would eliminate local
stability as a selection criterion, it would also imply that the
quantization is even more distant to that in the full theory where
such a quantization cannot be possible. Further clues as to the
necessary order of the evolution equation can come from studying Dirac
observables, which will be pursued elsewhere.

In section \ref{s:models}, we briefly describe the diagonal,
homogeneous models addressed in this paper. We recall the classical
framework and the corresponding loop quantization, including a
quantization of inverse triad components and general aspects of the
Hamiltonian constraint equation. Here we also discuss the two
criteria of admissibility of the continuum approximation and local stability.

In section \ref{hami}, we present a quantization of the Hamiltonian
constraint.  Since the Hamiltonian constraint now also has a potential
term depending on the spin connection, a quantization of the spin
connection is required. This is done using the quantization of inverse
triad components. We derive the (partial) difference equation
satisfied by the physical states and show that in the continuum
approximation, to the leading order, the approximating differential
equation is precisely the Wheeler-DeWitt equation with a specific
factor ordering dictated by the underlying loop quantization. We also
show that the requirement of local stability is satisfied. We
demonstrate that the singularity avoidance mechanism found in the case
of Bianchi I \cite{HomCosmo} continues to hold for these more general
models. We conclude that the proposed quantization of the Hamiltonian
constraint is satisfactory with respect to all our requirements.

In section \ref{bianchiIX}, we focus attention on the modification of
the potential implied by the quantization of the spin connection. The
non-trivial and non-perturbative behavior of the inverse triad
components may be expected to lead to substantial modifications of the
potential at small triad components. In view of the conjectured
central role of the Bianchi IX potential, we present a brief
description of this potential and its implications for the modified
approach to classical singularities.

\section{Diagonal Bianchi Class A Models}
\label{s:models}

In the following, we restrict attention to the so called diagonal,
Bianchi class A models. Among the class of homogeneous models i.e.
models whose symmetry group acts transitively on the spatial manifold,
is the sub-class of the Bianchi models for which the symmetry group
contains a subgroup with simply transitive action on the spatial
manifold.  The simply transitive subgroups are classified in terms of
three integers, $n^I$, parameterizing the structure constants as
\[
C^I_{JK}=\epsilon_{(I)JK}n^{I} \ .
\] 
Only class A models for which $C^I_{JI}=0$ admit a canonical
formulation.

In the connection formulation, Bianchi class A models are described in
terms of the invariant connections and densitized triads given by,
\[
 A^i_a ~ = ~ \phi^i_I \omega^I_a ~~,~~ E^a_i ~ = ~ p^I_i X^a_I\, 
\]
with constant and canonically conjugate $\phi^i_I$, $p^I_i$, where
$\omega^I$ are left-invariant (with respect to the subgroup acting
simply transitively) 1-forms on the spatial manifold $\Sigma$ and
$X_I$ are corresponding dual density-weighted vector fields. The
1-forms satisfy the Maurer-Cartan equations
\[
 \md\omega^I=-\Case{1}{2}C^I_{JK}\omega^J\wedge\omega^K\,.
\]

In the metric formulation, {\em diagonal} homogeneous models are those
for which the spatial metric can be written in a diagonal form such
that only three degrees of freedom remain.
In the connection formulation the situation is analogous. To
begin with one has nine degrees of freedom, $\phi^i_I$ but with a
three parameter gauge freedom of SU(2) rotations of the index $i$. A
further freedom, depending upon the parameters $n^I$, would have been
available if the left-invariant 1-forms were not thought of as a
background structure in parameterizing the invariant connections. To
get the same three degrees of freedom as in the diagonal metric
formulation, one then restricts the $\phi^i_I$ to a ``diagonal form",
$\phi^i_I := c_{(I)} \Lambda^i_I$ and correspondingly $p^I_i := p^I
\Lambda^{(I)}_i$ where the SO(3)-matrix $\Lambda$ includes gauge
degrees of freedom (see \cite{HomCosmo} for details). Such a
restriction specifies the diagonal models in the connection
formulation. This is {\it not} a symmetry reduction in the same sense
as restriction to isotropic models is. Nevertheless, these restricted
models can be analyzed by following procedures similar to those in the
context of symmetry reductions. As in reductions to isotropic models,
this leads to significant simplifications in such diagonalized models
because the gauge parameters in $\Lambda$ and (almost; see below)
gauge invariant parameters, $c_I$, are neatly separated
\cite{HomCosmo}.

\subsection{Classical Framework}
\label{ModClass}

The basic variables for diagonal Bianchi Class A models are specified
via
\begin{equation}
 A_a^i=c_{(I)}\Lambda_I^i\omega_a^I
\end{equation}
in terms of the `gauge invariant' coefficients $c_I$ with the pure
gauge degrees of freedom contained in the $SO(3)$-matrix
$\Lambda$. The components $c_I$ are not completely gauge invariant
but subject to residual discrete gauge transformations which change
the sign of two of the three components simultaneously. A diagonal
densitized triad has the form
\begin{equation}
 E_i^a=p^{(I)}\Lambda^I_iX_I^a
\end{equation}
where $X_I$ are left-invariant densitized vector fields dual to
$\omega^I$. Being an SO(3)-matrix, $\Lambda$ satisfies
\[
\Lambda^i_I \Lambda^J_i ~ = ~ \delta_I^J ~~ , ~~ 
\epsilon_{ijk} \Lambda^i_I \Lambda^j_J \Lambda^k_K ~ = ~ \epsilon_{IJK}
\]

The triad components $p^I$ are subject to the same residual gauge
transformations as the connection components $c_I$ to which they are
conjugate with basic Poisson bracket
\begin{equation}
 \{c_I,p^J\}=\gamma\kappa\delta_I^J
\end{equation}
where $\gamma$ is the Barbero--Immirzi parameter and $\kappa=8\pi G$
the gravitational constant. We will use the value
$\gamma=\frac{\log(2)}{\pi \sqrt{3}}\approx0.13$ fixed by the black
hole entropy calculations \cite{ABCK:LoopEntro,IHEntro}.

A diagonal co-triad has the form $e_a^i=a_{(I)}\Lambda_I^i\omega_a^I$
with
\begin{equation}
 p^1=|a_2a_3|\sgn(a_1)\quad,\quad p^2=|a_1a_3|\sgn(a_2) \quad,\quad
 p^3=|a_1a_2|\sgn(a_3)\,.
\end{equation}
Note that the components $p^I$ as well as $a_I$ can take negative
values, but only the overall sign $\sgn(p^1p^2p^3)$ (i.e., the
orientation) and the absolute values $|p^I|$ are gauge invariant. With
this form of the co-triad, we obtain, in fact, a diagonal metric
\[
 \md s^2=e_I^ie_J^i\omega^I\omega^J=\sum_Ia_I(\omega^I)^2\,.
\]

The extrinsic curvature also has diagonal form with components $K_I=-
\Case{1}{2}\dot{a}_I$ which appear in the relation
$c_I=\Gamma_I-\gamma K_I$ defining the Ashtekar connection components
$c_I$ in terms of the spin connection components $\Gamma_I$ and the
extrinsic curvature. The form of the spin connection introduces a
dependence of the framework on the particular model via the structure
constants \cite{HomCosmo}:
\begin{eqnarray} \label{SpinConn}
 \Gamma_I &=& \Case{1}{2}\left(\frac{a_J}{a_K}n^J+ \frac{a_K}{a_J}n^K-
   \frac{a_I^2}{a_Ja_K}n^I\right)\\
  &=& \Case{1}{2}\left(\frac{p^K}{p^J}n^J+ \frac{p^J}{p^K}n^K-
   \frac{p^Jp^K}{(p^I)^2}n^I\right) \mbox{ for $(I,J,K)$ an even
   permutation of $(1,2,3)$.}\nonumber
\end{eqnarray}
Note that in contrast to the full theory, the spin connection is a
covariant object within a homogeneous model since coordinate
transformations have to respect the symmetry. Consequently, if
non-zero, it cannot be made small by choosing appropriate local
coordinates. The only model which has identically vanishing spin
connection is the Bianchi I model with $n^I=0$. Otherwise, the spin
connection is non-zero and can even depend on the triad (it is a
constant $\Gamma=\frac{1}{2}$ in the closed isotropic model, a
submodel of Bianchi IX, with $n^I=1$ and $a_1=a_2=a_3$).

The structure constants also appear explicitly in the Hamiltonian
constraint \cite{HomCosmo}
\begin{eqnarray} \label{H}
 H &=& 2\kappa^{-1}\left\{\left[(c_2\Gamma_3+c_3\Gamma_2-\Gamma_2\Gamma_3)
     (1+\gamma^{-2})- n^1c_1-\gamma^{-2}c_2c_3\right]a_1\right.\nonumber\\
  &&+\left[(c_1\Gamma_3+c_3\Gamma_1-\Gamma_1\Gamma_3)
     (1+\gamma^{-2})- n^2c_2-\gamma^{-2}c_1c_3\right]a_2\nonumber\\
  &&\left.+\left[(c_1\Gamma_2+c_2\Gamma_1-\Gamma_1\Gamma_2)
     (1+\gamma^{-2})- n^3c_3-\gamma^{-2}c_1c_2\right]a_3\right\}\\
 &=& 2\kappa^{-1}\left[(\Gamma_2\Gamma_3-n^1\Gamma_1)a_1+
   (\Gamma_1\Gamma_3-n^2\Gamma_2)a_2
   +(\Gamma_1\Gamma_2-n^3\Gamma_3)a_3\right.\nonumber\\
  &&\left. -\Case{1}{4}(a_1\dot{a}_2\dot{a}_3+ a_2\dot{a}_1\dot{a}_3+
   a_3\dot{a}_1\dot{a}_2)\right]\,. \label{HH}
\end{eqnarray}
In order to derive the classical field equations (for which we
consider only positive $a_I$ and $p^I$) it is advantageous to
transform to new canonical variables
\begin{eqnarray}
 \pi_I &:=& 2K_Ip^{(I)}= - \dot{a}_Ia_I^{-1}a_1a_2a_3= - (\log a_I)'
 \quad\mbox{ and} \\ q^I &:=& \Case{1}{2}\log p^I
 \hspace{2cm}\mbox{such that}\hspace{2cm}\{q^I, \pi_J\} = \kappa
 \delta_I^J
\end{eqnarray}
where the prime denotes a derivative with respect to a new time
coordinate $\tau$ related to $t$ by $\md t=a_1a_2a_3\md\tau$
(corresponding to the lapse function $N=a_1a_2a_3$). With these new
variables we have $\{\pi_I,p^J\}= - 2\kappa p^{(I)}\delta_I^J$ and
\begin{eqnarray}
 \kappa NH &=& \kappa a_1a_2a_3H=
 2\left[p^1p^2(\Gamma_1\Gamma_2-n^3\Gamma_3)+
   p^1p^3(\Gamma_1\Gamma_3-n^2\Gamma_2)+
   p^2p^3(\Gamma_2\Gamma_3-n^1\Gamma_1)\right.\nonumber\\
   \label{ClassHam}
 && \left. - \Case{1}{4}(\pi_1\pi_2+\pi_1\pi_3+\pi_2\pi_3)\right]\\
 &=& \Case{1}{2}\left[(n^1)^2\left(\frac{p^2p^3}{p^1}\right)^2+
   (n^2)^2\left(\frac{p^1p^3}{p^2}\right)^2+
   (n^3)^2\left(\frac{p^1p^2}{p^3}\right)^2\right]\nonumber\\
 &&- n^1n^2(p^3)^2-
 n^1n^3(p^2)^2- n^2n^3(p^1)^2
 -\Case{1}{2}(\pi_1\pi_2+\pi_1\pi_3+\pi_2\pi_3)\\
 &=& \Case{1}{2}\left[(n^1)^2a_1^4+ (n^2)^2a_2^4+ (n^3)^2a_3^4\right]-
 n^1n^2a_1^2a_2^2- n^1n^3a_1^2a_3^2- n^2n^3a_2^2a_3^2\nonumber\\
 && -\Case{1}{2}(\pi_1\pi_2+\pi_1\pi_3+\pi_2\pi_3)\,.
\end{eqnarray}
Now one can easily derive the equations of motion,
\[
 (\log a_I)''= - \pi_I' \approx - \{\pi_I,a_1a_2a_3H\}
\]
which yield
\begin{eqnarray} \label{motion}
 \Case{1}{2}(\log a_1)'' &=& (n^2a_2^2-n^3a_3^2)^2-(n^1)^2a_1^4\nonumber\\
 \Case{1}{2}(\log a_2)'' &=& (n^1a_1^2-n^3a_3^2)^2-(n^2)^2a_2^4\\
 \Case{1}{2}(\log a_3)'' &=& (n^1a_1^2-n^2a_2^2)^2-(n^3)^2a_3^4\nonumber
\end{eqnarray}
or, using $(q^1)''= - \Case{1}{2}(\pi_2'+\pi_3')$
\[
 (q^I)''= -4a_I^2p^I\Gamma_In^I\,. 	
\]

For the Bianchi I model we have $n^I=0$ so that $\log
a_I=\alpha_I(\tau-\tau_{0,I})$. This implies the Kasner behavior
$a_I\propto t^{\alpha_I}$ where $t=e^\tau$ and the constraint requires
$0=\alpha_1\alpha_2+\alpha_1\alpha_3+\alpha_2\alpha_3=
\Case{1}{2}((\alpha_1+\alpha_2+\alpha_3)^2-
\alpha_1^2-\alpha_2^2-\alpha_3^2)$. The coefficients $\alpha_I$ can be
rescaled by choosing a different $t(\tau)$ which can be fixed by
requiring the conventional parameterization
$\alpha_1+\alpha_2+\alpha_3=1= \alpha_1^2+\alpha_2^2+\alpha_3^2$. As
usual, these equations can be solved only if one coefficient, say
$\alpha_1$, is negative and the other two are
positive. Correspondingly, one direction, the first, contracts whereas
the other two expand toward larger time. When we approach the
classical singularity at $t=0$, space shrinks only in two directions
while the third one expands unboundedly. The total volume however
continues to approach zero according to $a_1a_2a_3\propto
t^{\alpha_1+\alpha_2+\alpha_3}=t$. In \cite{HomCosmo} it has been
shown that the Kasner singularity disappears when the model is
quantized along the lines of loop quantum cosmology. For this result
it was important that quantum geometry is based on {\em densitized
triad} variables rather than co-triad variables. In those variables,
all $p^I\propto t^{1-\alpha_I}$ decrease to zero when one approaches
the classical singularity.
By contrast, for the
minisuperspace described in terms of the scale factors or co-triad
variables ($a_I$), the singularity is reached when at least one of
them goes to infinity. 
We paraphrase this by
saying that the classical singularity lies in the {\em interior} of
the triad minisuperspace (containing triads
of both orientations and also degenerate ones).

Bianchi models other than Bianchi I have non-vanishing structure
constants and thus the evolution of the three triad components can be
described as motion in a non-trivial potential given by
\begin{eqnarray} \label{potential}
 W(p^1,p^2,p^3) &=& 2\left\{p^1p^2(\Gamma_1\Gamma_2-n^3\Gamma_3)+
 p^1p^3(\Gamma_1\Gamma_3-n^2\Gamma_2)+
 p^2p^3(\Gamma_2\Gamma_3-n^1\Gamma_1)\right\}\nonumber\\
 &=& (p^1)^2\Gamma_1^2+(p^2)^2\Gamma_2^2+(p^3)^2\Gamma_3^2-
 (p^1\Gamma_1+p^2\Gamma_2+p^3\Gamma_3)^2 
\end{eqnarray}
which has infinite walls at small $p^I$ owing to the divergence of the
spin connection components (see Fig.~\ref{PotClass}). The evolution
can then be described approximately as a succession of Kasner epochs
with intermediate reflections at the potential \cite{Mixmaster}. For
the Bianchi IX model the reflections never stop and 
the classical evolution is believed to be chaotic \cite{Chaos}.

\subsection{Loop Quantization}
\label{s:Loop}

Diagonal homogeneous loop quantum cosmology \cite{HomCosmo} is first
formulated in the connection representation where an orthonormal basis
is given by the $\hat{p}^I$-eigenstates
\begin{equation}\label{n}
 |m_1,m_2,m_3\rangle:= |m_1\rangle\otimes |m_2\rangle\otimes
 |m_3\rangle
\end{equation}
with
\begin{equation} \label{cm}
 \langle c|m\rangle=\frac{\exp(\Case{1}{2}imc)}{\sqrt{2}\sin(\Case{1}{2}c)}\,.
\end{equation}
(The full Hilbert space of loop quantum cosmology is non-separable,
but for our purposes it is sufficient to use the separable subspace
spanned by the states (\ref{n}); see \cite{Bohr} for a detailed
discussion.)  The eigenvalues of the triad operators can be read off
from
\begin{equation}
 \hat{p}^I|m_1,m_2,m_3\rangle= \Case{1}{2}\gamma\lP^2m_I
 |m_1,m_2,m_3\rangle\,.
\end{equation}
Using the basic operators $\hat{p}^I$ one can define the volume
operator $\hat{V}=\sqrt{|\hat{p}^1\hat{p}^2\hat{p}^3|}$ which will be
used later. Its eigenstates are also $|m_1,m_2,m_3\rangle$ with
eigenvalues
\begin{equation} \label{V}
 V(m_1,m_2,m_3)=(\Case{1}{2}\gamma\lP^2)^{\frac{3}{2}}
 \sqrt{|m_1m_2m_3|}\,.
\end{equation}

A kinematical state $|s\rangle$ is described in the triad representation
by coefficients $s_{m_1, m_2, m_3}$ defined via,
\begin{equation}\label{TriadRep}
 |s\rangle=\sum_{m_1, m_2, m_3}s_{m_1, m_2, m_3}|m_1,m_2,m_3\rangle
\end{equation}
For a state to be gauge invariant under the residual gauge transformations, 
the coefficients $s_{m_1,m_2,m_3}$ have to satisfy
\begin{equation}
 s_{m_1,m_2,m_3}=s_{-m_1,-m_2,m_3}=s_{m_1,-m_2,-m_3}=
 s_{-m_1,m_2,-m_3}\,.
\end{equation}
These states are left invariant by the gauge invariant triad operators
$|\hat{p}^I|$ and the orientation operator
$\sgn(\hat{p}^1\hat{p}^2\hat{p}^3)$. In calculations it is often
easier to work with non-gauge invariant states in intermediate steps
and project to gauge invariant ones in the end.

Together with the basic derivative operators $\hat{p}^I$ we need
multiplication operators which usually arise from (point) holonomies
$h_I=\exp(c_{(I)}\Lambda_I^i\tau_i)=
\cos(\Case{1}{2}c_I)+2\Lambda_I^i\tau_i \sin(\Case{1}{2}c_I)$ with
action
\begin{eqnarray}
 \cos(\Case{1}{2}c_1) |m_1,m_2,m_3\rangle &=&
 \Case{1}{2}(|m_1+1,m_2,m_3\rangle+|m_1-1,m_2,m_3\rangle) \label{cos}\\ 
 \sin(\Case{1}{2}c_1) |m_1,m_2,m_3\rangle &=&
 -\Case{1}{2}i(|m_1+1,m_2,m_3\rangle- |m_1-1,m_2,m_3\rangle) \label{sin}
\end{eqnarray}
and correspondingly for $c_2$ and $c_3$.

\subsubsection{Inverse triad operators}

{}From the basic operators we can build more complicated ones. We will
later need a quantization of the spin connection which is a composite
operator containing several triad operators. In particular, it also
contains inverse powers of triad components which classically diverge
at the singularity. Since the triad operators have a discrete spectrum
containing zero, they do not have an inverse. However, general methods
of quantum geometry and loop quantum cosmology \cite{QSDV,InvScale}
imply that there exist well-defined operators quantizing inverse triad
components. To obtain these operators one makes use of a classical
reformulation, e.g.  
\[
 |p^1|^{-1}=4\gamma^{-2}\kappa^{-2}(\{c_1,\sqrt{|p^1|}\})^2=
 6\gamma^{-2}\kappa^{-2}
 \left(\tr\Lambda_1^i\tau_ih_I\{h_I^{-1},\sqrt{|p^1|}\}\right)^2
\]
which can then be quantized to
\begin{equation}\label{InvOpr}
\widehat{|p^1|^{-1}_j}=-36 \gamma^{-2}\lP^{-4}[j(j+1)(2j+1)]^{-2}
\left(\tr\nolimits_j\Lambda_1^i\tau_i
h_1[h_1^{-1},\sqrt{|\hat{p}^1|}]\right)^2\,.
\end{equation}
Here we have indicated that there are quantization ambiguities \cite{Ambig}
when one quantizes composite operators. The one, most relevant for this
paper, is indicated by the subscript $j$ of the trace and corresponds to 
the choice of representation while writing holonomies as multiplicative 
operators. 

This operator acts as
\begin{eqnarray} \label{InvOprAct}
\widehat{|p^I|_j^{-1}}|m_1, m_2, m_3\rangle & := & (\Case{1}{2}\gamma
\ell^2_p)^{
-1} 
{\cal{N}}_j^{-2} f_j( m_I) |m_1, m_2, m_3\rangle  ~~~\mbox{ where, } \\
f_j(m_I) & := & \left\{
\sum_{k = - j}^{j} k \sqrt{ | m_I + 2 k | } \right\}^2 ~~~ \mbox{ and } 
\nonumber \\
{\cal{N}}_j & := & \frac{j (j + 1) ( 2 j + 1 )}{3}\,. \nonumber
\end{eqnarray}
The discrete values $f_j(m)$ {\em decrease toward lower values\/} for
$m<2j$ \cite{Ambig}.  Thus, one can see that the classical
divergence of the inverse of $|p^I|$ at vanishing $p^I$ is explicitly
absent in the quantized operator. This will be seen to have further
consequences for the approach to the classical singularity.
Furthermore, the state $|m_1,m_2,m_3\rangle$ with $m_I=0$ (on which
the classical inverse triad would diverge) is actually annihilated by
$\widehat{|p^I|_j^{-1}}$ due to $f_j(0)=0$. This allows us to define
the inverse triad operator (not just its absolute value) by
$\widehat{(p^I)_j^{-1}} := \sgn(\hat{p}^I)
\widehat{|p^I|_j^{-1}}$ without ambiguity in defining the sign of
zero.  These inverse triad operators will be used later to find
well-defined operators quantizing the spin connection. Their
eigenvalues are
\begin{equation} \label{InvTriad}
 \widehat{(p^I)_j^{-1}}|m_1, m_2, m_3\rangle = (\Case{1}{2}\gamma
 \ell^2_p)^{-1} {\cal{N}}_j^{-2} \sgn(m_I) f_j( m_I) |m_1, m_2,
 m_3\rangle\,.
\end{equation}

\subsubsection{The Hamiltonian constraint}

Since the classical Hamiltonian constraint contains the structure
constants, it must have a model dependent quantization. The procedure
in the full theory \cite{QSDI} motivates to model the holonomy of a
closed loop by using the combination $h_Ih_Jh_I^{-1}h_J^{-1}$ in order
to quantize connection components in the constraint \cite{cosmoIII}.
This works in fact for the Bianchi I model, but it must be modified
for models with non-vanishing intrinsic curvature. In these models a
curve which is formed from four pieces of integral curves of vector
fields $X_I$ generating the symmetry action, which would have holonomy
$h_Ih_Jh_I^{-1}h_J^{-1}$, does not close to form a loop.  One needs a
correction which in \cite{cosmoIII} has been done by adding a
holonomy for the remaining curve multiplicatively. However this does
not result in an admissible evolution for the closed isotropic model
derived from Bianchi IX \cite{Stab} (an admissible quantization for
this model has recently been presented in \cite{Closed}). We thus have
to find a quantization which satisfies certain criteria for a
well-defined evolution.

To appreciate the criteria we recall that the general form of the
Hamiltonian constraint yields a {\em difference} equation for the wave
function in the triad representation because of the action of holonomy
operators \cite{cosmoIV}. This fact is directly related to the
discreteness of quantum geometry. However in the classical regime of large
volume and small extrinsic curvature, the geometry is the familiar
continuum (metric) geometry. Thus the difference equation must admit solutions
which can be very well approximated by solutions of {\em differential}
equations, namely the Wheeler--DeWitt equation which comes from the
(Schrodinger) quantization of continuum geometry. Such solutions are
slowly varying (in the classical regime) and can be interpolated by
slowly varying solutions of the Wheeler--DeWitt equation and are termed
as {\em pre-classical}. Since quantum geometry depends on the Barbero-Immirzi
parameter while the continuum geometry does not, absence of $\gamma$ dependence
in arriving at the differential approximation provides non-trivial 
constraints on the coefficients of the difference equation. Admissibility of 
{\em a continuum approximation} is our first criterion. 

However, admissibility of a continuum approximation alone is not enough.
Typically the order of the difference equation is larger than that of
the Wheeler--DeWitt equation. Consequently there are solutions of the
difference equation which are not interpolable by solutions of
differential equation. Therefore, if in some local region in the classical
regime we have a ``small" admixture of these extra solutions (small so
that the continuum approximation is valid), there is no guarantee that
contribution of the extra solutions remains small under subsequent
evolution, i.e. the continuum approximation may get invalidated. That
this does {\em not} happen is our second criterion. It is thus necessary
for the {\em stability} of the continuum approximation.

A detailed formulation of these criteria is given in \cite{Stab}. As a
first step, one divides the classical regime (large eigenvalues of the
triad operators) into smaller cells (local regions) which are still
large compared to the order of the difference equation. In each of these
cells, the coefficients of the difference equations are approximately
constant. This leads to approximating the difference equation by an
equation with constant coefficients. In the case of ordinary difference
equations that result in the isotropic models, the solutions are
controlled by the roots of a polynomial. The admissibility of continuum
approximation puts a set of conditions on these constant coefficients
which translate into conditions on the roots. For the isotropic case in
particular, the criterion requires existence of a root with value 1 and
multiplicity 2. The second criterion of local stability further requires
that all roots must have absolute value less than or equal to 1 in order
to prevent exponential growth of the non-pre-classical solutions. A further 
complex conjugation property of the coefficients of the difference equations
then implies that all roots in fact must have absolute value 1.

The quantization of the Hamiltonian constraint presented in the next section 
will be shown to satisfy both these criteria.

\section{Hamiltonian Constraints for Bianchi Class A Models}
\label{hami}

As noted in the introduction, models with non-vanishing intrinsic
curvature must have non-zero spin connection which, thanks to
homogeneity, cannot be made to vanish. Therefore, the Ashtekar
connection, a linear combination of the spin connection and the
extrinsic curvature, is not necessarily small even in a semiclassical
regime corresponding to small extrinsic curvature. While the
Hamiltonian constraint in quantum geometry is formulated naturally in
terms of the connection, for a semiclassical approximation it is
desirable to perform an expansion in the small extrinsic
curvature. This can only lead to the correct result if the spin
connection also appears explicitly in the constraint.  We will first
propose the quantized constraint operator, show how it reproduces the
classical expression and heuristically argue how it can admit a
continuum approximation with local stability. The properties of the
proposed operator will be elaborated in detail in the following
subsections.

The proposed Hamiltonian constraint operator is:
\begin{eqnarray}
\hat{H}  & = & 4 i ( \gamma \ell_p^2 \kappa )^{-1} 
\sum_{IJK} \epsilon^{IJK} \tr \left[ ~ \left\{ 
\gamma^{-2} \hat{h}_I(A, \Gamma) \hat{h}_J(A, \Gamma) 
(\hat{h}_I(A, \Gamma))^{-1}(\hat{h}_J(A, \Gamma))^{-1} \right.
\right. \nonumber \\
& & \hspace{4.5cm}
\left. \left. - 2 ( \hat{\Gamma}_I \hat{\Gamma}_J - n^L \hat{\Gamma}_L )
\Lambda_I \Lambda_J \right\} \left\{ h_K(A)
\left[ h_K^{-1}(A) , \hat{V} \right] \right\}~ \right] \label{Hop}
\end{eqnarray}
where
\begin{eqnarray}
\hat{h}_I(A, \Gamma) & := & e^{c_{(I)} \Lambda_I^i \tau_i} ~
e^{- \hat{\Gamma}_{(I)} \Lambda_I^i \tau_i} \label{HolonomyOpr}\\
\hat{\Gamma}_I & := & \frac{1}{2} \left[ 
\hat{p}^J n^K \widehat{(p^K)^{-1}} + 
\hat{p}^K n^J \widehat{(p^J)^{-1}} - 
\hat{p}^J \hat{p}^K n^I (\widehat{(p^I)^{-1}})^2 \right] \label{GammaOpr}
\end{eqnarray}
and the $\widehat{(p^I)^{-1}}$ are defined in
equation (\ref{InvTriad}). Since the inverse triad operators commute
with the triad operators, there are no ordering ambiguities in the
spin connection (\ref{GammaOpr}). Note that the holonomy operators
$h_K(A)=\exp(c_{(K)}\Lambda_K^i\tau_i)$ appearing in the second set of
braces do not need to contain the spin connection since it commutes
with $\hat{V}$ and would drop out.

In (\ref{Hop}) the quantized spin connection appears explicitly both
in the diagonal `potential' term and in holonomies $h_I(A, \Gamma)$.
For vanishing $\Gamma_I$ the constraint operator reproduces the one of
the Bianchi-I model \cite{cosmoIII,HomCosmo}. It is easy to see that
the term $(\Gamma_I\Gamma_J-n^L\Gamma_L) \Lambda_I\Lambda_J$ comes
from the curvature 2-form of the spin connection $\Gamma$ for a
Bianchi model. In fact, an expression via the curvature components
would be more general since it could also be used for Kantowski--Sachs
models.

The holonomies $h_I(A, \Gamma)$ are products of holonomies $h_I(A)$ and 
$(h_I(\Gamma))^{-1}$. Since we have to choose a quantization of the spin 
connection components $\Gamma_I$, which contain inverse triads, and also 
an ordering of the non-commuting $A$- and $\Gamma$-holonomies, there are 
many more quantization ambiguities compared to the Bianchi-I model.
However, qualitative effects can often be seen to be independent of such 
ambiguities.

To see that the proposed Hamiltonian operator goes over to the classical
expression in the classical limit, we can ignore the fact that connection 
components and the spin connection do not commute since this will only lead 
to effects of the order $\hbar$ which disappear in the classical limit. 
The holonomy product
$h_I(A)h_I(\Gamma)^{-1}$ can then be written formally as
\[
 h_I(A)h_I(\Gamma)^{-1}\sim \cos\left(\Case{1}{2}(c_I-\Gamma_I)\right)+
 2\Lambda_I^i\tau_i \sin\left(\Case{1}{2}(c_I-\Gamma_I)\right)
\]
which for small $K_I=-\gamma^{-1}(c_I-\Gamma_I)$ yields
\[
 h_I(A)h_I(\Gamma)^{-1}\sim 1-2\gamma\Lambda_I^i\tau_iK_I\,.
\]
On states (\ref{n}) with large labels $m_I$, the commutator containing
the volume operator turns into
\[
 h_K[h_K^{-1},\hat{V}]\sim -\Case{1}{2}i\gamma\lP^2 \Lambda_Ka_K
\]
so that the classical constraint in the form (\ref{HH}) is reproduced.

With regard to the continuum approximation and the local stability
conditions, we have to take into account the fact that the connection
does not commute with the spin connection because even small terms can
have significant effects should they lead to instabilities. The
products of holonomies then do not depend on the differences
$c_I-\Gamma_I$ only, as could be expected if they were to
commute. While ignoring the non-commutativity we have seen that
expressions like $\sin\left(\Case{1}{2}(c_I-\Gamma_I)\right)$ appear
which, however, are not well-defined in this form since there do not
exist $c_I$-operators in quantum geometry (only holonomies of the
Ashtekar connection are well-defined operators). Motivated by the form
of the holonomies in (\ref{HolonomyOpr}),
\begin{eqnarray*}
 h_I(A, \Gamma) &=& \left(\cos(\Case{1}{2}c_I)+
 2\Lambda_I\sin(\Case{1}{2}c_I)\right) \left(\cos(\Case{1}{2}\hat{\Gamma}_I)-
 2\Lambda_I\sin(\Case{1}{2}\hat{\Gamma}_I)\right)\\
 &=& \Case{1}{2}\left(e^{\frac{1}{2}ic_I}e^{-\frac{1}{2}i\hat{\Gamma}_I}+
 e^{-\frac{1}{2}ic_I}e^{\frac{1}{2}i\hat{\Gamma}_I}- 2i\Lambda_I
 (e^{\frac{1}{2}ic_I}e^{-\frac{1}{2}i\hat{\Gamma}_I}- 
 e^{-\frac{1}{2}ic_I}e^{\frac{1}{2}i\hat{\Gamma}_I})\right)\,,
\end{eqnarray*}
we define $\widehat{\sin}$ and $\widehat{\cos}$ operators 
by
\[
 h_I(A,\Gamma)=: \widehat{\cos}\left(\Case{1}{2}(c_I-\Gamma_I)\right)+ 
 2\Lambda_I \widehat{\sin}\left(\Case{1}{2}(c_I-\Gamma_I)\right)
\]
with action
\begin{eqnarray} \label{sinDiff}
 \widehat{\sin}\left(\Case{1}{2}(c_1-\Gamma_1)\right) |m_1, m_2, m_3\rangle 
 &:=& -\Case{1}{2}i \left(e^{\frac{1}{2}ic_1}e^{-\frac{1}{2}i\Gamma_1}-
 e^{-\frac{1}{2}ic_1}e^{\frac{1}{2}i\Gamma_1}\right) |m_1, m_2, m_3\rangle \\
 &=&
 -\Case{1}{2}i \left(e^{-\frac{1}{2}i\Gamma_1(m_I)} |m_1 + 1, m_2, m_3\rangle
 \right. \nonumber \\
 &  & ~~~~~~~~~~~~~~~ \left. -
 e^{\frac{1}{2}i\Gamma_1(m_I)} |m_1 -1, m_2, m_3\rangle\right) \nonumber
\end{eqnarray}
and analogously for $\widehat{\cos}$-operators. This means that, by
definition, the spin connection in trigonometric expressions of 
$c_I-\Gamma_I$, always acts first. In the final expression we used
eigenvalues $\Gamma_1(m_I)$ of the spin connection operators,
\[
 \Gamma_J|m_1,m_2,m_3\rangle:=\Gamma_J(m_I)|m_1,m_2,m_3\rangle\,.
\]

Operators of the form (\ref{sinDiff}) however, do not produce the basic
difference operator in the triad representation since they contain the
spin connection eigenvalues as coefficients. Nevertheless, for large
$m_I$ we have in the triad representation for a state $|s\rangle$
\begin{eqnarray*}
 \left(\widehat{\sin}\left(\Case{1}{2}(c_1-\Gamma_1)\right)
  s\right)_{m_1,m_2,m_3} &=&
 \Case{1}{2}i\left(e^{\frac{1}{2}i\Gamma_1(m_1+1,m_2,m_3)}
 s_{m_1+1,m_2,m_3} \right. \\
 & & ~~~~~~~~~~~~~~~ \left. - e^{-\frac{1}{2}i\Gamma_1(m_1-1,m_2,m_3)}
 s_{m_1-1,m_2,m_3}\right)\\
 &\sim& \Case{1}{2}i e^{-\frac{1}{2}im_1\Gamma_1(m_I)}
 (\tilde{s}_{m_1+1,m_2,m_3}- \tilde{s}_{m_1-1,m_2,m_3})
\end{eqnarray*}
where 
\begin{equation} \label{stilde}
\tilde{s}:=\exp(i\hat{p}^I\hat{\Gamma}_I/\gamma\lP^2)s\,,
\end{equation}
i.e.,
\begin{equation}
 \tilde{s}_{m_1,m_2,m_3}=
 \exp\left(\Case{1}{2}i\sum_Im_I\Gamma_I(m_J)\right)s_{m_1,m_2,m_3}\,.
\end{equation}
For large $m_I$, therefore, the operators (\ref{sinDiff}) do act as
basic difference operators on the wave function $\tilde{s}$ up to
phase factors, and analogously operators
$\widehat{\cos}\left(\Case{1}{2}(c_I-\Gamma_I)\right)$ act as basic
mean operators.

Using the wave function $\tilde{s}$ instead of $s$, the constraint
operator (\ref{Hop}) contains, besides diagonal (potential) terms, a sum of
products of a basic difference operator and a basic mean operator for
two of the three discrete parameters $m_I$. The resulting difference
equation for $\tilde{s}_{m_1,m_2,m_3}$ will thus be of the same form as the
difference equation of the Bianchi-I model. Consequently, the quantization 
(\ref{Hop}) can be expected to yield a locally stable evolution equation
as in the case of Bianchi-I. We will discuss this in detail in the
following subsections. 

\subsection{Discrete evolution equation}

We now return to the constraint operator (\ref{Hop}) and obtain the
difference equation in detail. From the definitions it follows that,
\begin{eqnarray}
\hat{h}_I(A, \Gamma) & = & 
\frac{e^{\frac{i}{2} c_I} e^{- \frac{i}{2} \hat{\Gamma}_I} +
e^{- \frac{i}{2} c_I} e^{\frac{i}{2} \hat{\Gamma}_I}}{2} + 2 \Lambda_I~
\frac{e^{\frac{i}{2} c_I} e^{- \frac{i}{2} \hat{\Gamma}_I} -
e^{- \frac{i}{2} c_I} e^{\frac{i}{2} \hat{\Gamma}_I}}{2 i} \nonumber \\
& =: & \hat{C}_I - 2 i \Lambda_I \hat{S}_I \label{holonomy}\\
\left( \hat{h}_I(A, \Gamma) \right)^{-1} & = & 
\frac{e^{\frac{i}{2} \hat{\Gamma}_I} e^{- \frac{i}{2} c_I} +
e^{- \frac{i}{2} \hat{\Gamma}_I} e^{\frac{i}{2} c_I}}{2} + 2 \Lambda_I~
\frac{e^{\frac{i}{2} \hat{\Gamma}_I} e^{- \frac{i}{2} c_I} -
e^{- \frac{i}{2} \hat{\Gamma}_I} e^{\frac{i}{2} c_I}}{2 i} \nonumber \\
& =: & \hat{C}_I^{\dagger} - 2 i \Lambda_I \hat{S}_I^{\dagger}
\label{inv-holonomy}\\
\hat{h}_K(A) \left[ \hat{h}^{-1}_K(A) , \hat{V} \right] & = & 
\left\{ \hat{V} - \cos(\Case{1}{2}c_K) \hat{V} \cos(\Case{1}{2}c_K)
- \sin(\Case{1}{2}c_K) \hat{V} \sin(\Case{1}{2}c_K) \right\} 
\nonumber \\
& & - 2 \Lambda_K \left\{ \sin(\Case{1}{2}c_K) \hat{V} \cos(\Case{1}{2}c_K)
- \cos(\Case{1}{2}c_K) \hat{V} \sin(\Case{1}{2}c_K) \right\}
\end{eqnarray}

Note that the connection components $c_I$ do not have a `hat' on them as
these do not exist as operators. Their exponentials however are well
defined, multiplicative operators. For notational convenience these
`hats' have been suppressed. The `sin', $\hat{S}$, has been defined 
{\em without} the factor of $i$ also for future notational convenience.

Then the trace in (\ref{Hop}) gives (suppressing the hats on the
operators):
\begin{eqnarray}
& & 2 \gamma^{-2}\left\{ C_I C_J C^{\dagger}_I C^{\dagger}_J + S_I C_J
  S^{\dagger}_I C^{\dagger}_J +
C_I S_J C^{\dagger}_I S^{\dagger}_J - S_I S_J S^{\dagger}_I
  S^{\dagger}_J \right\} \nonumber \\
& & \hspace{3.5cm}
\times\left\{ \hat{V} - \cos(\Case{1}{2}c_K) \hat{V} \cos(\Case{1}{2}c_K)
- \sin(\Case{1}{2}c_K) \hat{V} \sin(\Case{1}{2}c_K) \right\} \nonumber \\
& & -2 \gamma^{-2} ~ \epsilon_{IJK} ~
 \left\{ S_I S_J C^{\dagger}_I C^{\dagger}_J - C_I S_J S^{\dagger}_I
  C^{\dagger}_J + 
S_I C_J C^{\dagger}_I S^{\dagger}_J + C_I C_J S^{\dagger}_I
  S^{\dagger}_J \right\} \nonumber \\
& & \hspace{3.5cm} 
\times\left\{ \sin(\Case{1}{2}c_K) \hat{V} \cos(\Case{1}{2}c_K)
- \cos(\Case{1}{2}c_K) \hat{V} \sin(\Case{1}{2}c_K) \right\}
\nonumber \\
& &- \epsilon_{IJK} ~\left\{ \Gamma_I \Gamma_J - n^K \Gamma_K \right\} 
 \left\{ \sin(\Case{1}{2}c_K) \hat{V} \cos(\Case{1}{2}c_K)
- \cos(\Case{1}{2}c_K) \hat{V} \sin(\Case{1}{2}c_K) \right\}
\end{eqnarray}

Doing the sum over $IJK$ finally yields,
\begin{eqnarray}
\left( \frac{ \kappa \gamma^3 \ell^2_p}{8 i} \right) \hat{H} & = &
\left\{ 
C_1 C_2 C^{\dagger}_1 C^{\dagger}_2 + S_1 C_2 S^{\dagger}_1 C^{\dagger}_2 +
C_1 S_2 C^{\dagger}_1 S^{\dagger}_2 - S_1 S_2 S^{\dagger}_1 S^{\dagger}_2 
\right. \nonumber \\
& & \left. 
 - C_2 C_1 C^{\dagger}_2 C^{\dagger}_1 - S_2 C_1 S^{\dagger}_2 C^{\dagger}_1 -
C_2 S_1 C^{\dagger}_2 S^{\dagger}_1 + S_2 S_1 S^{\dagger}_2
S^{\dagger}_1 \right\} \nonumber \\
& & \hspace{1.5cm} \times\left\{ \hat{V} - \cos(\Case{1}{2}c_3) \hat{V}
\cos(\Case{1}{2}c_3)
- \sin(\Case{1}{2}c_3) \hat{V} \sin(\Case{1}{2}c_3) \right\}
\nonumber \\ 
& &  -\left\{ 
S_1 S_2 C^{\dagger}_1 C^{\dagger}_2 - C_1 S_2 S^{\dagger}_1 C^{\dagger}_2 +
S_1 C_2 C^{\dagger}_1 S^{\dagger}_2 + C_1 C_2 S^{\dagger}_1 S^{\dagger}_2 
\right. \nonumber \\
& & \left.
+ S_2 S_1 C^{\dagger}_2 C^{\dagger}_1 - C_2 S_1 S^{\dagger}_2 C^{\dagger}_1 +
S_2 C_1 C^{\dagger}_2 S^{\dagger}_1 + C_2 C_1 S^{\dagger}_2 S^{\dagger}_1 
\right\} 
\nonumber \\
& & \hspace{1.5cm} 
\times\left\{ \sin(\Case{1}{2}c_3) \hat{V} \cos(\Case{1}{2}c_3)
- \cos(\Case{1}{2}c_3) \hat{V} \sin(\Case{1}{2}c_3) \right\}
 \nonumber \\
& &  -\gamma^2 ~ \left\{ \Gamma_1 \Gamma_3 - n^3 \Gamma_3 \right\} 
 \left\{ \sin(\Case{1}{2}c_3) \hat{V} \cos(\Case{1}{2}c_3)
- \cos(\Case{1}{2}c_3) \hat{V} \sin(\Case{1}{2}c_3) \right\}
 \nonumber \\
& & \mbox{ $ + $ cyclic} 
\end{eqnarray}

The terms containing $\hat{V}$ and $\hat{\Gamma}_I$ are both diagonal
when acting on the basis vectors $|m_1, m_2, m_3\rangle$. They give:
\begin{eqnarray}
\hat{\Gamma}_I |m_1, m_2, m_3\rangle & = & \Gamma_I(m_1, m_2, m_3) |m_1,
m_2, m_3\rangle ~~~~~ \mbox{where,} \\
\Gamma_I(m_1, m_2, m_3) & := & \frac{1}{2} {\cal{N}}_j^{-2}  \left( m_J
n^K f_j(m_K) + m_K n^J f_j(m_J) - m_J m_K n^I {\cal{N}}_j^{-2}
f_j^2(m_I) \right) \nonumber 
\end{eqnarray}
using the notation of (\ref{InvOprAct}).

Next,

$\left( \hat{V} - \cos(\Case{1}{2}c_3) \hat{V} \cos(\Case{1}{2}c_3)
- \sin(\Case{1}{2}c_3) \hat{V} \sin(\Case{1}{2}c_3) \right)
  |m_1, m_2, m_3\rangle ~ = $
\begin{equation}
\left( V(m_1, m_2, m_3) - \frac{1}{2} V(m_1, m_2, m_3 + 1) - 
\frac{1}{2} V(m_1, m_2, m_3 - 1) \right) |m_1, m_2, m_3\rangle 
\end{equation}

$ \left( \sin(\Case{1}{2}c_3) \hat{V} \cos(\Case{1}{2}c_3)
- \cos(\Case{1}{2}c_3) \hat{V} \sin(\Case{1}{2}c_3) \right)
|m_1, m_2, m_3\rangle ~ = $
\begin{equation}
\frac{i}{2}\left( V(m_1, m_2, m_3 + 1) - V(m_1, m_2, m_3 - 1) \right)
|m_1, m_2, m_3\rangle 
\end{equation}

The action of the $\hat{C}_I, \hat{S}_I, \hat{C}^{\dagger}_I,
\hat{S}^{\dagger}_I$ on the basis states can be expressed as:
\begin{eqnarray}
\hat{C}_1 |m_1, m_2, m_3\rangle & = & \;\frac{1}{2} \sum_{\epsilon_1 = 
\pm 1} e^{ - \frac{i}{2} \epsilon_1 \Gamma_1(m_1, m_2, m_3) }
|m_1 + \epsilon_1, m_2, m_3\rangle \\
\hat{S}_1 |m_1, m_2, m_3\rangle & = & \;\frac{1}{2} \sum_{\epsilon_1 = 
\pm 1} \epsilon_1 ~ e^{ - \frac{i}{2} \epsilon_1 \Gamma_1(m_1, m_2, m_3) }
|m_1 + \epsilon_1, m_2, m_3\rangle \\
\hat{C}^{\dagger}_1 |m_1, m_2, m_3\rangle & = & \;\frac{1}{2}
\sum_{\epsilon_1^{\prime} = 
\pm 1} e^{ - \frac{i}{2} \epsilon_1^{\prime} \Gamma_1(m_1 +
\epsilon_1^{\prime}, m_2, m_3) }
|m_1 + \epsilon_1^{\prime}, m_2, m_3\rangle \\
\hat{S}^{\dagger}_1 |m_1, m_2, m_3\rangle & = & -\frac{1}{2}
\sum_{\epsilon_1^{\prime} = 
\pm 1}\epsilon_1^{\prime} ~ e^{ - \frac{i}{2} \epsilon_1^{\prime}
\Gamma_1(m_1 + \epsilon_1^{\prime}, m_2, m_3) }
|m_1 + \epsilon_1^{\prime}, m_2, m_3\rangle 
\end{eqnarray}

In each group of four terms, the phase factor will be the same since each
consists of two $\hat{C}, \hat{S}$ type of terms and two of
$\hat{C}^{\dagger}, \hat{S}^{\dagger}$ types of terms. We will just
have different factors of the $\epsilon_1, \epsilon_1^{\prime}$. Thus,
for example,
$$
\left\{ 
C_1 C_2 C^{\dagger}_1 C^{\dagger}_2 + S_1 C_2 S^{\dagger}_1 C^{\dagger}_2 +
C_1 S_2 C^{\dagger}_1 S^{\dagger}_2 - S_1 S_2 S^{\dagger}_1 S^{\dagger}_2 
\right\} |m_1, m_2, m_3\rangle  ~ = $$
$$
\frac{1}{16} \sum_{\epsilon_1, \epsilon_2, \epsilon_1^{\prime},
\epsilon_2^{\prime}} e^{- \frac{i}{2} \left\{ 
( \epsilon_1 + \epsilon_1^{\prime} ) 
\Gamma_1(m_1 + \epsilon_1^{\prime}, m_2 + \epsilon_2^{\prime}, m_3) 
~+~\epsilon_2
\Gamma_2(m_1 + \epsilon_1^{\prime}, m_2 + \epsilon_2^{\prime}, m_3) 
~+~\epsilon_2^{\prime} 
\Gamma_2(m_1, m_2 + \epsilon_2^{\prime}, m_3) \right\} } $$
\begin{equation}
~~~\times\left( 1 - \epsilon_1 \epsilon_1^{\prime} - \epsilon_2
\epsilon_2^{\prime} - \epsilon_1 \epsilon_1^{\prime} \epsilon_2
\epsilon_2^{\prime} \right) ~~ |m_1 + \epsilon_1 + \epsilon_1^{\prime}, 
m_2 + \epsilon_2 + \epsilon_2^{\prime}, m_3\rangle \,,
\end{equation}
$$
\left\{ 
C_2 C_1 C^{\dagger}_2 C^{\dagger}_1 + S_2 C_1 S^{\dagger}_2 C^{\dagger}_1 +
C_2 S_1 C^{\dagger}_2 S^{\dagger}_1 - S_2 S_1 S^{\dagger}_2 S^{\dagger}_1 
\right\} |m_1, m_2, m_3\rangle  ~ = $$
$$
\frac{1}{16} \sum_{\epsilon_1, \epsilon_2, \epsilon_1^{\prime},
\epsilon_2^{\prime}} e^{- \frac{i}{2} \left\{ 
( \epsilon_2 + \epsilon_2^{\prime} ) 
\Gamma_2(m_1 + \epsilon_1^{\prime}, m_2 + \epsilon_2^{\prime}, m_3) 
~+~\epsilon_1
\Gamma_1(m_1 + \epsilon_1^{\prime}, m_2 + \epsilon_2^{\prime}, m_3) 
~+~\epsilon_1^{\prime} 
\Gamma_1(m_1 + \epsilon_1^{\prime}, m_2, m_3) \right\} } $$
\begin{equation}
~~~\times\left( 1 - \epsilon_1 \epsilon_1^{\prime} - \epsilon_2
\epsilon_2^{\prime} - \epsilon_1 \epsilon_1^{\prime} \epsilon_2
\epsilon_2^{\prime} \right) ~~ |m_1 + \epsilon_1 + \epsilon_1^{\prime}, 
m_2 + \epsilon_2 + \epsilon_2^{\prime}, m_3\rangle \,,
\end{equation}
$$
\left\{ 
S_1 S_2 C^{\dagger}_1 C^{\dagger}_2 - C_1 S_2 S^{\dagger}_1 C^{\dagger}_2 +
S_1 C_2 C^{\dagger}_1 S^{\dagger}_2 + C_1 C_2 S^{\dagger}_1 S^{\dagger}_2 
\right\} |m_1, m_2, m_3\rangle  ~ = $$
$$
\frac{1}{16} \sum_{\epsilon_1, \epsilon_2, \epsilon_1^{\prime},
\epsilon_2^{\prime}} e^{- \frac{i}{2} \left\{ 
( \epsilon_1 + \epsilon_1^{\prime} ) 
\Gamma_1(m_1 + \epsilon_1^{\prime}, m_2 + \epsilon_2^{\prime}, m_3) 
~+~\epsilon_2
\Gamma_2(m_1 + \epsilon_1^{\prime}, m_2 + \epsilon_2^{\prime}, m_3) 
~+~\epsilon_2^{\prime} 
\Gamma_2(m_1, m_2 + \epsilon_2^{\prime}, m_3) \right\} } $$
\begin{equation}
~~~\times\left( 
\epsilon_1 \epsilon_2 + \epsilon_1^{\prime} \epsilon_2 - 
\epsilon_1 \epsilon_2^{\prime} + \epsilon_1^{\prime} \epsilon_2^{\prime} 
\right) ~~ |m_1 + \epsilon_1 + \epsilon_1^{\prime}, 
m_2 + \epsilon_2 + \epsilon_2^{\prime}, m_3\rangle
\end{equation}
and
$$
\left\{ 
S_2 S_1 C^{\dagger}_2 C^{\dagger}_1 - C_2 S_1 S^{\dagger}_2 C^{\dagger}_1 +
S_2 C_1 C^{\dagger}_2 S^{\dagger}_1 + C_2 C_1 S^{\dagger}_2 S^{\dagger}_1 
\right\} |m_1, m_2, m_3\rangle  ~ = $$
$$
\frac{1}{16} \sum_{\epsilon_1, \epsilon_2, \epsilon_1^{\prime},
\epsilon_2^{\prime}} e^{- \frac{i}{2} \left\{ 
( \epsilon_2 + \epsilon_2^{\prime} ) 
\Gamma_2(m_1 + \epsilon_1^{\prime}, m_2 + \epsilon_2^{\prime}, m_3) 
~+~\epsilon_1
\Gamma_1(m_1 + \epsilon_1^{\prime}, m_2 + \epsilon_2^{\prime}, m_3) 
~+~\epsilon_1^{\prime} 
\Gamma_1(m_1 + \epsilon_1^{\prime}, m_2, m_3) \right\} } $$
\begin{equation}
~~~\times\left( 
\epsilon_1 \epsilon_2 - \epsilon_1^{\prime} \epsilon_2 + 
\epsilon_1 \epsilon_2^{\prime} + \epsilon_1^{\prime} \epsilon_2^{\prime} 
\right) ~~ |m_1 + \epsilon_1 + \epsilon_1^{\prime}, 
m_2 + \epsilon_2 + \epsilon_2^{\prime}, m_3\rangle \,.
\end{equation}

It is convenient to define different basis vectors at this stage,
namely,
\begin{eqnarray}
|m_1, m_2, m_3\rangle 
& := & 
e^{ \frac{1}{2}i \sum_I m_I \Gamma_I(m_1, m_2, m_3)}
\widetilde{|m_1, m_2, m_3\rangle}
\end{eqnarray}
corresponding to the transformation (\ref{stilde}). This introduces
additional phases which will turn out to be useful in our discussion
of the continuum approximation. The transformation to the triad
representation via,
\begin{equation}
|s\rangle := \sum_{m_1, m_2, m_3} \tilde{s}_{m_1, m_2, m_3} \widetilde{
|m_1, m_2, m_3 \rangle},
\end{equation}
leads to an equation for the $\tilde{s}_{m_1, m_2, m_3}$ : 
\begin{eqnarray}
0 & = & 
\sum_{\epsilon_1, \epsilon_2, \epsilon_1^{\prime}, \epsilon_2^{\prime}} 
\gamma^{-2} A_{12}(m_1, m_2, m_3; \epsilon_1, \epsilon_2, \epsilon_1^{\prime},
\epsilon_2^{\prime}) 
\tilde{s}_{ m_1 - \epsilon_1 - \epsilon_1^{\prime}, 
m_2 - \epsilon_2 - \epsilon_2^{\prime}, m_3} \nonumber \\
& & 
+\sum_{\epsilon_2, \epsilon_3, \epsilon_2^{\prime}, \epsilon_3^{\prime}} 
\gamma^{-2} A_{23}(m_1, m_2, m_3; \epsilon_2, \epsilon_3, \epsilon_2^{\prime},
\epsilon_3^{\prime}) 
\tilde{s}_{ m_1, m_2 - \epsilon_2 - \epsilon_2^{\prime}, 
m_3 - \epsilon_3 - \epsilon_3^{\prime} } \nonumber \\
& & 
+\sum_{\epsilon_3, \epsilon_1, \epsilon_3^{\prime}, \epsilon_1^{\prime}} 
\gamma^{-2} A_{31}(m_1, m_2, m_3; \epsilon_3, \epsilon_1, \epsilon_3^{\prime},
\epsilon_1^{\prime}) 
\tilde{s}_{ m_1 - \epsilon_1 - \epsilon_1^{\prime}, m_2,
m_3 - \epsilon_3 - \epsilon_3^{\prime} } \nonumber \\
& & \hspace{1.5cm} + ~ B(m_1, m_2, m_3) \tilde{s}_{m_1, m_2, m_3} ~~,
\label{DiffEqn}
\end{eqnarray}
where the coefficients are rather complicated expressions given by:
\begin{eqnarray}
{v}^{\pm}_{12}(m_1, m_2, m_3; \epsilon_1, \epsilon_2, \epsilon_1^{\prime},
\epsilon_2^{\prime}) & := & 
V(m_1 - \epsilon_1 - \epsilon_1^{\prime},
m_2 - \epsilon_2 - \epsilon_2^{\prime}, m_3 + 1) \nonumber \\
& & \pm ~ V(m_1 - \epsilon_1 - \epsilon_1^{\prime},
m_2 - \epsilon_2 - \epsilon_2^{\prime}, m_3 - 1)\,, \\
{u}_{12}(m_1, m_2, m_3; \epsilon_1, \epsilon_2, \epsilon_1^{\prime},
\epsilon_2^{\prime}) & := & 
V(m_1 - \epsilon_1 - \epsilon_1^{\prime},
m_2 - \epsilon_2 - \epsilon_2^{\prime}, m_3 ) \nonumber \\
& & - ~ \frac{1}{2} {v}^+_{12}(m_1, m_2, m_3; \epsilon_1, \epsilon_2, 
\epsilon_1^{\prime}, \epsilon_2^{\prime}) \,,
\end{eqnarray}
\begin{eqnarray}
\phi_{12,1}(m_1, m_2, m_3; \epsilon_1, \epsilon_2, \epsilon_1^{\prime},
\epsilon_2^{\prime}) & := & - \sum_{I} m_I \Gamma_I(m_1, m_2, m_3)
\nonumber \\
& & + ~ (\epsilon_1 + \epsilon_1^{\prime}) \Gamma_1(m_1 - \epsilon_1, m_2 -
\epsilon_2, m_3) \nonumber \\
& & + ~ (\epsilon_2 + \epsilon_2^{\prime}) \Gamma_2(m_1 - \epsilon_1, m_2 -
\epsilon_2, m_3) \nonumber \\
& & + ~ \epsilon_1^{\prime} \left\{~ 
\Gamma_1(m_1 - \epsilon_1, m_2 - \epsilon_2 - \epsilon_2^{\prime},
m_3) \right. \nonumber \\
& & \left. ~~~~~ - ~ \Gamma_1(m_1 - \epsilon_1, m_2 - \epsilon_2, m_3) ~\right\}
\end{eqnarray}
and
\begin{eqnarray}
\phi_{12,2}(m_1, m_2, m_3; \epsilon_1, \epsilon_2, \epsilon_1^{\prime},
\epsilon_2^{\prime}) & := & - \sum_{I} m_I \Gamma_I(m_1, m_2, m_3)
\nonumber \\
& & + ~ (\epsilon_1 + \epsilon_1^{\prime}) \Gamma_1(m_1 - \epsilon_1, m_2 -
\epsilon_2, m_3) \nonumber \\
& & + ~ (\epsilon_2 + \epsilon_2^{\prime}) \Gamma_2(m_1 - \epsilon_1, m_2 -
\epsilon_2, m_3) \nonumber \\
& & + ~ \epsilon_2^{\prime} \left\{~ 
\Gamma_2(m_1 - \epsilon_1 - \epsilon_1^{\prime},
m_2 - \epsilon_2, m_3) \right. \nonumber \\
& & \left. ~~~~~ - ~ \Gamma_2(m_1 - \epsilon_1, m_2 - \epsilon_2, m_3)
~\right\} 
\end{eqnarray}

With obvious abbreviations, e.g., $(m_1, m_2, m_3) \equiv \vec{m}$, 
$(\epsilon_1, \epsilon_2, \epsilon^{\prime}_1, \epsilon^{\prime}_2)
\equiv \vec{\epsilon}_{12}$ and \\
$~~~~~A_{12}(m_1, m_2, m_3; \epsilon_1, \epsilon_2, \epsilon_1^{\prime},
\epsilon_2^{\prime}) ~ \equiv ~ A_{12}(\vec{m}; \ \vec{\epsilon}_{12})
$ etc. , we have
\begin{eqnarray}
A_{12}(\vec{m}; \ \vec{\epsilon}_{12}) & := & i u_{12}(\vec{m}; 
\vec{\epsilon}_{12}) \, \left( 1 - \epsilon_1 \epsilon_1^{\prime} - \epsilon_2
\epsilon_2^{\prime} - \epsilon_1 \epsilon_1^{\prime} \epsilon_2
\epsilon_2^{\prime} \right) 
\left( e^{- \frac{i}{2} \phi_{12,2}(\vec{m};\, \vec{\epsilon}_{12}) } -
e^{- \frac{i}{2} \phi_{12,1}(\vec{m};\, \vec{\epsilon}_{12}) }\right)
\nonumber \\
& & \hspace{0.5cm} + ~ \frac{1}{2} v^-_{12}(\vec{m};\, \vec{\epsilon}_{12}) 
\left\{ 
e^{- \frac{i}{2} \phi_{12,2}(\vec{m};\, \vec{\epsilon}_{12}) } 
\left(\epsilon_1 \epsilon_2 + \epsilon_1^{\prime} \epsilon_2 - 
\epsilon_1 \epsilon_2^{\prime} + \epsilon_1^{\prime} \epsilon_2^{\prime} 
\right) \right. \nonumber \\
& & \hspace{3.5cm} +\left.  e^{- \frac{i}{2}
\phi_{12,1}(\vec{m};\, \vec{\epsilon}_{12}) } 
\left(\epsilon_1 \epsilon_2 - \epsilon_1^{\prime} \epsilon_2 + 
\epsilon_1 \epsilon_2^{\prime} + \epsilon_1^{\prime} \epsilon_2^{\prime} 
\right) \right\}
\end{eqnarray}
and
\begin{eqnarray}
B(\vec{m}) & := & 8 \left( e^{\frac{i}{2} \sum_I m_I
\Gamma_I(\vec{m})} \right) \nonumber \\
& & \times\left[v^-_{12}(\vec{m}; \vec{0}) \left\{
\Gamma_1(\vec{m}) \Gamma_2(\vec{m}) - n^3 \Gamma_3(\vec{m}) \right\} \right.
\nonumber \\
& & v^-_{23}(\vec{m}; \vec{0}) \left\{
\Gamma_2(\vec{m}) \Gamma_3(\vec{m}) - n^1 \Gamma_1(\vec{m}) \right\} 
\nonumber \\
& & v^-_{31}(\vec{m}; \vec{0}) \left. \left\{
\Gamma_3(\vec{m}) \Gamma_1(\vec{m}) - n^2 \Gamma_2(\vec{m}) \right\} 
\frac{}{} \right]\nonumber \\
& & - 2 \kappa \gamma \ell_p^2 \left( e^{\frac{i}{2} \sum_I m_I
\Gamma_I(\vec{m}) } \right) \hat{H}_{m_1, m_2, m_3}^{\text{matter}}
\end{eqnarray}

The matter Hamiltonian acting on the matter field dependence of the
wave function has also been incorporated in the coefficient
$B(\vec{m})$.

Noting that the coefficients $u_{12}, A_{12}, B$ involve $v^{\pm}_{12}, V$ and
that the volume eigenvalues have a separable product structure, we can
absorb the volume factors in the wave functions as follows.
Define,
\begin{eqnarray}
d^{\pm}(m) & :=  & \left\{\begin{array}{cl} 
\sqrt{1 + m^{-1}} \pm \sqrt{1 - m^{-1}}  &  \mbox{ \hspace{1.0cm} if
  $m \ne 0$} \\
0 & \mbox{ \hspace{1.0cm} if $ m = 0$ }
\end{array} \right.
\end{eqnarray}
Then, 
\begin{eqnarray}
v^{\pm}_{12}(\vec{m}; \vec{\epsilon}) & = & 
V(m_1 - \epsilon_1 - \epsilon_1^{\prime}, 
m_2 - \epsilon_2 - \epsilon_2^{\prime},  m_3) d^{\pm}( m_3 ) \\
u_{12}(\vec{m}; \vec{\epsilon}) & = & 
V(m_1 - \epsilon_1 - \epsilon_1^{\prime}, 
m_2 - \epsilon_2 - \epsilon_2^{\prime},  m_3) \left\{ 1 - \frac{d^+( m_3
)}{2} \right\} 
\end{eqnarray}

Now defining $\tilde{t}_{m_1, m_2, m_3} := V(\vec{m}) \tilde{s}_{m_1,
m_2, m_3}$ and $\hat{\rho}^{\text{matter}}_{m_1, m_2, m_3} :=
\hat{H}^{\text{matter}}_{m_1, m_2, m_3}/V(\vec{m}) $ for $V(m_1, m_2,
m_3) \neq 0$, one sees that all volume eigenvalue factors are absorbed
in the wave function. This preserves the gauge invariance condition on
the $\tilde{S}$ but now in addition, due to the explicit volume
eigenvalues, we also have that $\tilde{t}_{m_1, m_2, m_3} = 0$ if any
of the $m_i$'s equal zero.

For Bianchi I, the spin connection is zero. The phases $\phi_{12,1},
\phi_{12,2}, \ldots $ vanish. The coefficient $B$ reduces to the
matter term only, while $A_{12}$ reduces to :
\begin{equation}
A_{12}(\vec{m}; \vec{\epsilon}_{12}) ~ = ~ 
 d^-(m_3) \left(\epsilon_1 \epsilon_2 + \epsilon_1^{\prime}
 \epsilon_2^{\prime} \right) 
\end{equation}
Equation (\ref{DiffEqn}) then becomes, 
\begin{eqnarray}
& & ~~~d^-(m_1)\left( \tilde{t}_{m_1, m_2 + 2, m_3 + 2} 
+ \tilde{t}_{m_1, m_2 - 2, m_3 -2} 
- \tilde{t}_{m_1, m_2 + 2, m_3 - 2} 
- \tilde{t}_{m_1, m_2 - 2, m_3 + 2} \right)  \nonumber \\
& & + ~ d^-(m_2)\left( \tilde{t}_{m_1 + 2, m_2, m_3 + 2} 
+ \tilde{t}_{m_1 - 2, m_2, m_3 - 2} 
- \tilde{t}_{m_1 - 2, m_2, m_3 + 2} 
- \tilde{t}_{m_1 + 2, m_2, m_3 - 2} \right)  \nonumber \\
& & + ~ d^-(m_3)\left( \tilde{t}_{m_1 + 2, m_2 + 2, m_3} 
+ \tilde{t}_{m_1 - 2, m_2 - 2, m_3} 
- \tilde{t}_{m_1 + 2, m_2 - 2, m_3} 
- \tilde{t}_{m_1 - 2, m_2 + 2, m_3} \right)  \nonumber \\
& & ~~ = ~ -2 \kappa \gamma^3 \ell_p^2 \hat{\rho}_{m_1, m_2,
m_3}^{\text{matter}} \tilde{t}_{m_1, m_2, m_3}
\end{eqnarray}
which matches with the equation derived in \cite{HomCosmo}.

\subsection{Continuum approximation}

We will first derive a simplified difference equation which
approximates the exact one for large $m_I$. Then using an interpolating
function, we will show that the approximate difference
equation gives a differential equation for the interpolating function,
which matches with the Wheeler--DeWitt equation at the level of
leading terms. This will verify that the continuum approximation is
admissible.

Consider first $m_I \gg 1$. Then, for $k \ll m$, $f_j(m + k) \rightarrow 
\frac{{\cal{N}}_j^2}{m}$ which is {\em independent} of $k$ to the leading
order. This implies that 
\begin{equation}
\Gamma_I ~ \sim ~ \frac{1}{2}\left\{ n^K \frac{m_J}{m_K} + n^J
\frac{m_K}{m_J} - n^I \frac{m_J m_K}{m_I^2} \right\} . \label{GamAsy}
\end{equation}

The phases $\phi_{12,1}, \phi_{12,2}$ etc.\ all become equal to the common
phase appearing in the coefficient $B$ and cancel out. The coefficients
$A_{IJ}$ become exactly the same as in Bianchi-I with $d^-(m_I) \approx
m_I^{-1}$ and the equation becomes,
\begin{eqnarray} \label{Asy}
- 2 \kappa \gamma \ell_p^2 \hat{\rho}^{\text{matter}}_{m_1, m_2, m_3}
~ \tilde{t}_{ m_1, m_2, m_3} 
& =  & \gamma^{-2} m_3^{-1} \sum_{\epsilon_1, \epsilon_2, \epsilon_1^{\prime},
\epsilon_2^{\prime}} 
(\epsilon_1 \epsilon_2 + \epsilon_1^{\prime} \epsilon_2^{\prime})
\tilde{t}_{ m_1 - \epsilon_1 - \epsilon_1^{\prime}, 
m_2 - \epsilon_2 - \epsilon_2^{\prime}, m_3} \nonumber \\
& & ~~~ + ~ 8 ~ m_3^{-1}  
\left( \Gamma_1 \Gamma_2 - n^3 \Gamma_3 \right) \tilde{t}_{ m_1, m_2,
m_3} 
~ + ~ \mbox{cyclic} 
\end{eqnarray}

To arrive at a differential equation, we look for a function
$\tilde{T}(p^1, p^2, p^3)$ such that $\tilde{t}_{m_1, m_2, m_3} :=
\tilde{T}( p^1(m_1), p^2(m_2), p^3(m_3))$ satisfies the approximate
difference equation.  Here, $p^I(m_I) := \frac{1}{2}\gamma \ell_p^2
m_I$ and thus $\delta p^I = \frac{1}{2}\gamma \ell_p^2$ is the change
induced when the $m_I$ change by 1.  In particular, this implies:
\begin{eqnarray}
\tilde{t}_{m_1 + k_1, m_2 + k_2, m_3 + k_3} & := & \tilde{T}(p^I(m_I +
k_I)) ~ = ~ \tilde{T}( p^I(m_I) + k_I \delta p^I ) \nonumber \\
& \approx & \tilde{T}(p^I(m_I)) + \sum_I k_I \delta p^I \frac{\partial
\tilde{T}}{\partial p^I} + \frac{1}{2} \sum_{IJ} k_I k_J \delta p^I
\delta p^J \frac{\partial^2 \tilde{T}}{\partial p^I \partial p^J}
\nonumber \\
& & ~~~~ + ~~~~ \cdots
\end{eqnarray}

It follows that terms of order $(\delta p^I)^0, (\delta p^I)^1$ cancel
out leaving a second order differential expression to the leading
order. We also have the pre-factors $\gamma^{-2} m_I^{-1}$ which can be
eliminated using $m_I = \frac{2 p^I}{\ell_p^2}\gamma^{-1}$. The spin connection
then becomes just the classical expression without any factors of $\gamma 
\ell_p^2$.  {\em All} factors of $\gamma$ cancel out. Multiplying by 
$\frac{1}{8}p^1 p^2 p^3$ the final equation becomes:
$$ \left[ \frac{\ell_p^4}{4}p^1 p^2 \frac{\partial^2 \tilde{T}(p^1,
p^2, p^3)}{\partial p^1 \partial p^2} + p^1 p^2 (\Gamma_1 \Gamma_2 -
n^3 \Gamma_3) \tilde{T}(p^1, p^2, p^3) \right] ~ + ~ \mbox{cyclic}
$$
\begin{equation} \label{wdw}
\hspace{3.0cm} = ~ -\frac{1}{2} \kappa\,p^1 p^2 p^3
\hat{\rho}^{\text{matter}}(p^1,p^2,p^3) \tilde{T}(p^1, p^2, p^3)
\end{equation}

We have thus verified that the constraint operator (\ref{Hop}), admits a
continuum approximation. 

Note that using an interpolating function, $\tilde{T}(p^1, p^2, p^3)$,
one can always arrive at some differential expression. That {\em all}
dependence on $\gamma$ disappears from the differential equation is a
non-trivial property of the locally approximated difference equation
and is essential for matching with the Wheeler--DeWitt equation which
knows nothing about the Barbero-Immirzi parameter. Had the phases not
canceled, we would have acquired terms from the Taylor expansion of
the $\Gamma_I$, which could have produced lower order differentials
with a $\gamma$-dependence left over, destroying the hope of matching
with the Wheeler--DeWitt equation. That the spin connection controls
the phases in a cancelable manner is thus also a non-trivial property
of the constraint operator. Finally, the fact that we get a purely
second order differential expressions is determined by the structure
of the coefficients in the locally approximated difference equation
\cite{Stab}. This is directly responsible for getting the
Wheeler-DeWitt equation in a {\em particular} factor ordering
appearing in (\ref{wdw}).

The wave function $\tilde{T}$ has been obtained after some
redefinitions.  Since $V(\vec{m}) \rightarrow \sqrt{|p^1 p^2 p^3|}$ in
the pre-classical limit, $\tilde{T}$ is related to $\tilde{S}$ as
$\tilde{T} \sim \sqrt{|p^1 p^2 p^3|} ~ \tilde{S}$. Thus, the
Wheeler--DeWitt equation for our original pre-classical wave function
$\tilde{S}$ is
\[
\left[ \frac{\ell_p^4}{4}p^1 p^2 \frac{\partial^2 \sqrt{|p^1
p^2 p^3|} ~ \tilde{S}(p^1, p^2, p^3)}{\partial 
p^1 \partial p^2} + p^1 p^2 (\Gamma_1 \Gamma_2 - n^3 \Gamma_3)
\sqrt{|p^1 p^2 p^3|} ~ \tilde{S}(p^1, p^2, p^3) \right] 
~ + ~ \mbox{cyclic} 
\]
\begin{equation}
\hspace{3.0cm} = ~ -\frac{1}{2} \kappa\,|p^1 p^2 p^3|^{\frac{3}{2}}
       \hat{\rho}^{\text{matter}}(p^1,p^2, p^3)\tilde{S}(p^1, p^2, p^3) \,.
\end{equation}
Note that the ordering of an isotropic sub-model is not determined
uniquely by using $p^1=p^2=p^3=:p$ because $p^1$ and
$\partial/\partial p^2$, say, commute in the anisotropic model but not
after using isotropy. However, there is also a unique ordering for
isotropic models \cite{IsoCosmo,SemiClass} which can be derived {\em
after} doing the reduction in loop quantum cosmology.

\subsection{Local Stability}
Our next requirement is about local stability of the evolution
equation around pre-classical solutions.  This is most directly
formulated and illustrated  in the context of an {\em ordinary} difference
equation \cite{Stab}. The equation we have however is a {\em partial} difference
equation. Although, we do get a partial difference equation with
constant coefficients in a local approximation, their solutions are
not as easily analyzable in terms of roots of some
polynomial. Nevertheless, the property of local stability can still be
analyzed.

Since we have three $m_I$'s, even if all are taken to be large, it is
possible to have a subset of them to be much larger compared to the
remaining ones.  To begin with, we consider a neighborhood of a point
where all $m_I$ are large and of the same order; other regions can be
explored similarly. For $\gamma$ we assume the value derived from
black hole entropy calculations which is of order $10^{-1}$. Since all
$m_I$'s are large, it follows exactly as before, that $\Gamma_I$ are
given by the equation (\ref{GamAsy}) and all phases cancel out giving
the locally approximated difference equation (\ref{Asy}). Furthermore,
since we assume the $m_I$ to be about the same size, the $\Gamma_I$
are approximately equal and taking the Bianchi IX model for
definiteness, $\Gamma_I\approx\frac{1}{2}$. We obtain
\begin{eqnarray}
& & ~~~d^-(m_1)\left( \tilde{t}_{m_1, m_2 + 2, m_3 + 2} 
+ \tilde{t}_{m_1, m_2 - 2, m_3 -2} 
- \tilde{t}_{m_1, m_2 + 2, m_3 - 2} 
- \tilde{t}_{m_1, m_2 - 2, m_3 + 2} \right)  \nonumber \\
& & + ~ d^-(m_2)\left( \tilde{t}_{m_1 + 2, m_2, m_3 + 2} 
+ \tilde{t}_{m_1 - 2, m_2, m_3 - 2} 
- \tilde{t}_{m_1 - 2, m_2, m_3 + 2} 
- \tilde{t}_{m_1 + 2, m_2, m_3 - 2} \right)  \nonumber \\
& & + ~ d^-(m_3)\left( \tilde{t}_{m_1 + 2, m_2 + 2, m_3} 
+ \tilde{t}_{m_1 - 2, m_2 - 2, m_3} 
- \tilde{t}_{m_1 - 2, m_2 + 2, m_3} 
- \tilde{t}_{m_1 + 2, m_2 - 2, m_3} \right)  \nonumber \\
& & ~~~ = -2\kappa\gamma^3\lP^2\hat{\rho}^{\text{matter}}_{m_1,m_2,m_3}
\tilde{t}_{m_1,m_2,m_3}+
2\gamma^2(d^-(m_1)+d^-(m_2)+d^-(m_3))\tilde{t}_{m_1,m_2,m_3}\,.
\end{eqnarray}

In this equation, $m_1, m_2$ and $m_3$ all play the same role, but
without loss of generality we can select $m_3$ to be the evolution
parameter in terms of which we analyze local stability. To express it
clearly, let us introduce the notation
\[
z^{\nu_1, \nu_2}_{m_3} := \tilde{t}_{m_1 + \nu_1, m_2 + \nu_2, m_3},
~~~ \nu_1, \nu_2 = 0, \pm 2
\]
which allows us to write the equation as
\begin{eqnarray}
&&d^-(m_1) (z^{0, 2}_{m_3 + 2} - z^{0, 2}_{m_3 - 2}) - 
d^-(m_1) (z^{0, -2}_{m_3 + 2} - z^{0, -2}_{m_3 - 2}) \nonumber \\
&& ~~~~~~~~~~~~~~~ + d^-(m_2) (z^{2, 0}_{m_3 + 2} - z^{2, 0}_{m_3 - 2}) - 
d^-(m_2) (z^{-2, 0}_{m_3 + 2} - z^{-2, 0}_{m_3 - 2}) \nonumber\\
&& = -d^-(m_3)(z^{2,2}_{m_3}+ z^{-2,2}_{m_3}- z^{2,-2}_{m_3}-
z^{-2,-2}_{m_3}) \nonumber \\
&& ~~~~~~~~~~~~~~~ - 2\gamma^2 (\gamma P+d^-(m_1)+d^-(m_2)+d^-(m_3))
z^{0,0}_{m_3}
\end{eqnarray}
where
\[
 P:=\kappa\lP^2\rho^{\text{matter}}
\]
is a small constant (in the local approximation) characterizing the
size of the matter contribution. 

For $m_I$'s of the same order $m$, the equation can be further simplified as
\begin{eqnarray} \label{ApproxEqn}
&&(z^{0, 2}_{m_3 + 2} - z^{0, 2}_{m_3 - 2}) - 
(z^{0, -2}_{m_3 + 2} - z^{0, -2}_{m_3 - 2}) +
(z^{2, 0}_{m_3 + 2} - z^{2, 0}_{m_3 - 2}) - 
(z^{-2, 0}_{m_3 + 2} - z^{-2, 0}_{m_3 - 2})\nonumber\\
& \approx & -(z^{2,2}_{m_3}+ z^{-2,2}_{m_3}- z^{2,-2}_{m_3}-
z^{-2,-2}_{m_3})- 2\gamma^2 (3+\gamma m P) z^{0,0}_{m_3} \label{Stab}
\end{eqnarray}

We are interested in checking if perturbation of slowly varying (along
all three directions) solutions, continue to remain as perturbations
under the $m_3$-evolution. To
this end, we assume that we have some fixed solution which is slowly
varying. Then the first bracket on the right hand side is small while
the second one is suppressed by $\gamma^2 \sim 10^{-2}$ ($mP$ is
small even for large $m$ because the energy density decreases at least
as $m^{-3}$). 

This is a linear equation among four combinations analogous to the
equation of a plane in $\mathbb{R}^4$ passing close to the origin. It
can be solved approximately by introducing three, fixed orthonormal
vectors, $\vec{{\cal E}}^a, a = 1, 2, 3$, each of which is orthogonal
to the vector $(1,-1,1,-1)$ in $\mathbb{R}^4$. Explicitly,
\begin{eqnarray}
(z^{0, 2}_{m_3 + 2} - z^{0, 2}_{m_3 - 2}) 
& \approx & \sum_{a = 1}^3 Q_a(m_3) ({\cal{E}}^a )_{0, 2} \nonumber \\
(z^{0, -2}_{m_3 + 2} - z^{0, -2}_{m_3 - 2}) 
& \approx & \sum_{a = 1}^3 Q_a(m_3) ({\cal{E}}^a )_{0, -2} \nonumber \\
(z^{2, 0}_{m_3 + 2} - z^{2, 0}_{m_3 - 2}) 
& \approx & \sum_{a = 1}^3 Q_a(m_3) ({\cal{E}}^a )_{2, 0} \nonumber \\
(z^{-2, 0}_{m_3 + 2} - z^{-2, 0}_{m_3 - 2}) 
& \approx & \sum_{a = 1}^3 Q_a(m_3) ({\cal{E}}^a )_{-2, 0}
\end{eqnarray}
In the above, corrections of the order
$z_{m_3}^{\nu_1,\nu_2}-z_{m_3}^{\nu_1', \nu_2'}$ as well as of the
order $\gamma^2 (3+\gamma m P) z^{0,0}_{m_3}$ are ignored and
$Q_a(m_3)$ are three functions of $m_3$. The $Q_a(m_3)$ can be assumed 
to be small for $z$ to represent a solution close to a pre-classical one.

Thus we obtain ordinary, {\em non-homogeneous} difference equations with
constant coefficients for the four $z^{\nu_1, \nu_2}_{m_3}$. All the
equations have the same structure, and their solutions consist of a
combination of the particular solution and linear combinations of 
solutions of the corresponding homogeneous equations. The {\em
difference} between a perturbed solution and the fixed solution of
course satisfies the {\em homogeneous} equations.

The local stability now requires that these differences do
not grow exponentially. Equivalently, the characteristic roots for the
homogeneous equations must have absolute value less than or equal to one.

In our case, the characteristic polynomials are the same for all
four equations namely, $z^4 - 1=0$, with roots equal to $\pm 1, \pm i$.
These obviously satisfy the condition of local stability.

Because of local stability, perturbations of slowly varying solutions
continue to remain close to them under evolution along the $m_3$ direction. 
This also ensures {\em self-consistency} when we consider evolutions
along the $m_1, m_2$ directions in a similar manner by keeping the
corresponding right hand sides close to zero.

Had we not neglected the terms on the right hand side of eq.
(\ref{ApproxEqn}), we would have got an extra, common non-homogeneous
term which is 1/2 of the right hand side of eq.(\ref{ApproxEqn}).
Clearly, this does not affect the behavior of the differences.

To summarize, although we got a partial difference equation, we were able
to implement the idea behind the local stability criterion by obtaining
a system of ordinary difference equations. That these equations are
non-homogeneous, does not affect the stability properties \cite{Stab}. 

\subsection{Absence of Singularities}

We have now verified that the proposed quantization (\ref{Hop}) has a
valid continuum approximation and is locally stable around
pre-classical solutions. Thus, it is possible to start with initial
data at large volume which are close to a continuum solution and to
follow the evolution to smaller volume, toward the classical
singularity. In this regime, quantum effects of the discreteness and
also ordering issues become more and more important and the full
equation has to be considered without approximations. Fortunately, it
is possible to see essential features of the absence of singularities
quite generally in a way similar to the isotropic case \cite{Sing}.
Since the constraint equation is used as a recurrence relation for the
wave function, the term with lowest order in the labels $m_I$ should
have non-vanishing coefficient, except when one of the labels vanishes
such that the corresponding component of the wave function, i.e.\ its
value at singular surfaces in minisuperspace, will drop out (see
\cite{HomCosmo} for details). It can readily be verified that this
general behavior is true for the evolution equation of an arbitrary
Bianchi class A model. The difference equation resulting from
(\ref{Hop}) is more complicated than in the isotropic \cite{IsoCosmo}
or Bianchi I \cite{HomCosmo} case; but since the exponentiated spin
connection never vanishes and the additional potential term containing
the spin connection is diagonal, only the commutator of holonomies
with the volume operator determines whether or not a lowest order
coefficient vanishes. The structure resulting from this term is the
same as in the Bianchi I case, and thus the evolution is
singularity-free in the same way.

This shows that the wave function can be extended uniquely from the
region of minisuperspace with positive orientation to the region with
negative orientation and that the orientation again provides us with a
branch `before' the classical singularity. 

For Bianchi models other than Bianchi I, however, the {\em classical
approach to the singularity} is more complicated than the simpler
Kasner behavior.  This is most dramatic in the Bianchi IX model where
different Kasner epochs succeed each other infinitely many times. It
turns out, as we will discuss in the next section, that the classical
approach to singularities is modified at small scales by quantum
geometry effects which is expected to lead to a simpler picture.

\subsection{Quantization Choices}

Even after imposing the condition of local stability the constraint
expression is far from being unique. There may be other choices for
the kinetic term which lead to a locally stable behavior, but they
would have to be very special and difficult to find. The advantage of
the quantization presented here is that it is rather simple and
provides a general formula for all diagonal Bianchi class A models at
once without having to find an expression in each case separately.

As for factor ordering ambiguities, the main issue is that of ordering
the commutator to the right or to the left. This ordering is the same
here as in the Bianchi I case and also in the full theory. Thus, the
special issues one has to deal with in models with non-vanishing
intrinsic curvature do not change this result. Since the non-singular
evolution depends significantly on this kind of ordering, it is
robust. The main purpose of this paper is to prove the existence of
one quantization for all models in our class which has the correct
semiclassical limit, results in a locally stable evolution and is also
non-singular in the quantum regime. That such a general quantization
fulfilling all the different requirements exists is a non-trivial fact
and provides a further check of the methods employed in the full
theory.

The second issue, besides the singularity-free evolution, which is
robust under the choices we have is the semiclassical behavior and
corrections to it. The main effects in this context are insensitive to
factor ordering and other choices. As we will discuss in the next
section, there are non-perturbative effects coming from the potential
term which lead to drastic differences to the classical
behavior. Since they come from the potential term and are based on
general features of quantizations of inverse triad components, they
have to be present in any quantization.

\section{Modified Bianchi IX Behavior}
\label{bianchiIX}

The quantization (\ref{Hop}) of the Hamiltonian constraint yields
consistent quantizations of the dynamics of Bianchi class A models
which in the limit of large triad components reproduce the correct
classical behavior. At finite values of $p^I$, however, there are
always correction terms to the classical equations of motion
(\ref{motion}) which can change the classical evolution. It is
possible to derive different types of corrections and to include them
in effective classical equations of motion. There are the following
sources of correction terms: First, we can keep more terms beyond the
leading order in a Taylor expansion of difference operators which
results in a higher order Wheeler--DeWitt equation. These correction
terms would correspond to higher curvature terms in an effective
action. The second source is the non-commutativity of the $A$-holonomy
$h_I(A)$ and the $\Gamma$-holonomy $h_I(\Gamma)$ in (\ref{Hop}). Since
we have, for example, $\{c_1,\Gamma_1\}= \gamma\kappa
p^2p^3(p^1)^{-3}n^1$ which in a quantization would appear as the
diagonal operator
\[
 i\hbar\Lambda_1\widehat{\{c_1,\Gamma_1\}}=
 -h_1(A)[h_1(A)^{-1},\hat{\Gamma}_1] \ ,
\]
the coefficients of the Wheeler--DeWitt equation receive corrections
of the order $o(\lP^2/p^I)$. Thirdly, we have to use well-defined
inverse metric operators when we quantize the spin connection
components in (\ref{Hop}) which leads to a modified behavior at small
scales (as in the isotropic case where it implies inflation
\cite{Inflation}).

Since the last effect is non-perturbative while the first two are
perturbative (in $\gamma$), it can be expected to have the most
dramatic consequences. Therefore, we will focus on this type of
correction in our discussion of a modified approach to classical
singularities. As in \cite{Inflation}, we exploit the fact that there
are quantization ambiguities \cite{Gaul} which can extend the range of
a modified classical behavior \cite{Ambig}. If we choose the ambiguity
parameter large enough it is reasonable to use effective classical
equations of motion to describe the trajectory of a quantum wave
packet while still experiencing the modified behavior at small
scales. In particular, the spin connection components will always
decrease when we approach the classical singularity close enough
whereas they always have some diverging components in the purely
classical scenario. The diverging components lead to infinitely high
walls in the potential (\ref{potential}) on minisuperspace and to an
infinite number of reflections in the Bianchi IX model leading to
chaotic behavior \cite{Chaos}. With the modified spin connection, the walls 
do not grow arbitrarily high but instead shrink once a small triad component
is reached. Consequently, one expects that the reflections will cease after a
finite number of times and the wave packet will simply move through
the classical singularity in a final Kasner epoch.

Here, we will make this explicit by deriving the potential and its
changes due to the modified spin connection.  A detailed analysis of
the modified approach to the singularity will be presented elsewhere
\cite{BianchiIX, BianchiIXDetail}.

\begin{figure}
 \begin{center}
 \includegraphics[width=12cm,height=8cm]{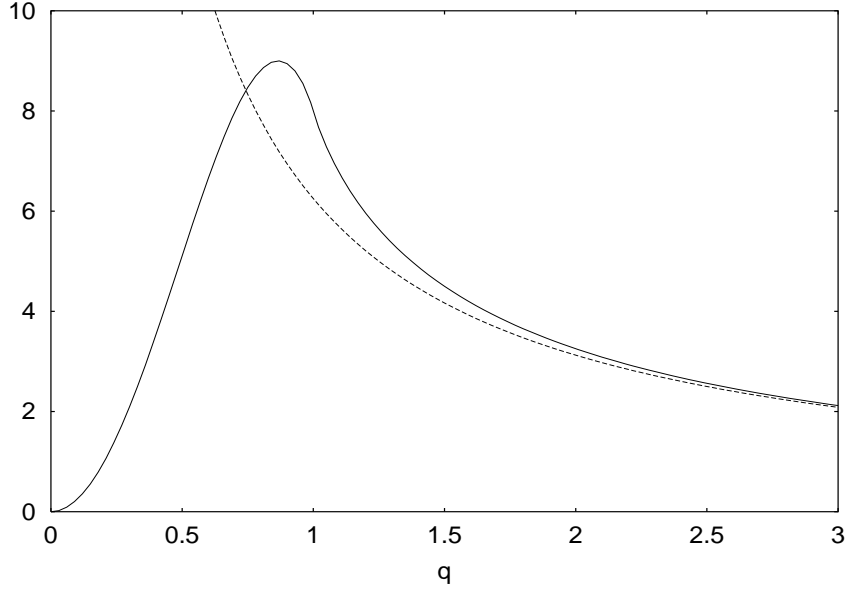}
 \end{center}
 \caption{The function $F(q)$ (solid line) compared to its large-$q$
 limit $q^{-1}$ (dashed).
 \label{F}}
\end{figure}

The quantized spin connection operators $\hat{\Gamma}_I$ have been
defined in equation (\ref{GammaOpr}) with the inverse triad operators
defined in equation (\ref{InvOprAct}). The functions $f_j(m)$ can
be approximated (with an accuracy increasing with $j$) in a manner
similar to that discussed in \cite{Ambig}. In terms of $q :=
\frac{m}{2 j}$, one obtains
\begin{eqnarray}
f_j(2jq) & ~~ \approx ~~ & \frac{2}{9} j^5 F(q)
\mbox{\hspace{5.5cm}where,} \nonumber \\
F(q) & := & \frac{4}{25}\left\{
3 \left[ (q + 1)^{\frac{5}{2}} - |q - 1|^{\frac{5}{2}} \right] - 5 q 
\left[ (q + 1)^{\frac{3}{2}} - {\mathrm{sgn}}(q - 1) |q -
1|^{\frac{3}{2}} \right]\right\}^2\\
&\sim& q^{-1} ~~~~\!\mbox{ for }q\gg1\nonumber\\
&\sim& 4q^2 ~~~~\mbox{ for }q\ll1\nonumber
\end{eqnarray}
(see Fig.\ \ref{F}).
Limiting behaviors for the eigenvalues of $\widehat{(p^I)_{j}^{-1}}$
are
\begin{eqnarray}
(p^I)^{-1}_{j}(m_J) & ~ \approx ~ & \left(
  \gamma \ell_p^2 j \right)^{-1}  F(m_I/2 j) \sgn(m_I)\\
& \sim & p^I(m_J)^{-1} 
\mbox{\hspace{4.7cm}for $|p^I(m_J)| \gg \gamma \lP^2 j$ } \\
& \sim & 4 \left(\gamma \lP^2j\right)^{-3} 
p^I(m_J)^2 \sgn(m_I)
\mbox{\hspace{1.5cm}for $|p^I(m_J)| \ll \gamma \lP^2 j$ }
\end{eqnarray}

Here $p^I(m_J) = \frac{1}{2} \gamma \ell_p^2 m_I$ are the eigenvalues of
$\hat{p}^I$. The eigenvalues of the spin connection operators are
obtained as,
\begin{eqnarray}
\Gamma^{(j)}_I(\vec{m}) & \approx & \frac{1}{4 j}\left[ ~~~~ n^J m_K
  F(m_J/2j) \sgn(m_J) + n^K
m_J F(m_K/2j) \sgn(m_K) \right. \nonumber\\
 &&  \hspace{0.8cm} - \left. n^I m_J m_K F^2(m_I/2j)/2j~~~~\right] \\
&& \nonumber \\
& \sim & \frac{1}{2} \left[ n^J \frac{p^K(m_K)}{p^J(m_J)} + n^K
\frac{p^J(m_J)}{p^K(m_K)}- n^I \frac{p^J(m_J) p^K( m_K)}{(p^I)^2(m_I)} \right]
\mbox{\hspace{1.3cm} $\left(\frac{|p^I|}{\gamma \lP^2 j} \gg 1 \right)$} \\ 
&& \nonumber \\
& \sim & \frac{1}{4 j^3} \left[ \, n^J m_K m_J^2 \sgn(m_J) + n^K
m_J m_K^2 \sgn(m_K) - n^I \frac{m_J m_K m_I^4}{2 j^3} \, \right]
\nonumber \\
& \approx & 2 \left(\gamma \ell_p^2 j\right)^{-3} 
p^J(m_J) p^K(m_K) \left[ ~ n^J |p^J(m_J)| + n^K |p^K(m_K)| ~ 
\right] 
\mbox{\hspace{0.4cm} $\left(\frac{|p^I|}{\gamma \lP^2 j} \ll 1 \right)$} 
\end{eqnarray}

In the last equation we have dropped the sub-leading third term.
Evidently, the eigenvalues of the spin connection operators approach
their classical expressions for large values of the triad components
while for small values the quantum modification shows up. To find
the potential in this regime, we must use the first line of eqn.
(\ref{potential}) since the second follows by using the classical
expression for $\Gamma_I$. For $|p^I|\ll\gamma\lP^2j$ this leads to,
\begin{eqnarray} 
W_j(\vec{p}) & := & 2 p^1 p^2 (\Gamma^{(j)}_1 \Gamma^{(j)}_2
- n^3 \Gamma^{(j)}_3)
~ + ~ \mbox{cyclic} \nonumber \\
& \approx &  \left[\,8 (\gamma \ell_p^2 j)^{-6} (p^1 p^2 p^3)^2 (n^2
  |p^2| + n^3 |p^3|)(n^1 |p^1|+ n^3 |p^3|)) \right.\nonumber \\
& & \left. - 4 (\gamma \ell_p^2 j)^{-3} (p^1 p^2)^2 n^3 (n^1 |p^1| +
  n^2 |p^2| ) 
\, \right] ~ + ~ \mbox{cyclic} \ .
\end{eqnarray}

Dropping the $o(j^{-6})$ term compared to the $o(j^{-3})$, for $|p^I|
\ll \gamma \ell_p^2 j$, we get
\begin{eqnarray} \label{NearPot}
W_j(\vec{p}) & \approx & - 4 (\gamma \ell_p^2 j)^{-3} \left[~~
n^1 n^2 (p^3)^2 \ (\ |p^1|^3 + |p^2|^3\ )\  + \ 
n^2 n^3 (p^1)^2 \ (\ |p^2|^3 + |p^3|^3\ )\  \right. \nonumber \\
&& \hspace{2.0cm} \left. + ~ 
n^3 n^1 (p^2)^2 \ (\ |p^3|^3 + |p^1|^3\ ) ~~ \right]
\end{eqnarray}

By contrast, for $|p^I| \gg \gamma \ell_p^2 j$, the potential takes the
form,
\begin{eqnarray} \label{FarPot}
W_j(\vec{p}) & \approx & \frac{1}{2} \left[~
\left\{ p^1 p^2 (p^3)^{-1} n^3\ \right\}^2 + 
\left\{ p^2 p^3 (p^1)^{-1} n^1\ \right\}^2 + 
\left\{ p^3 p^1 (p^2)^{-1} n^2\ \right\}^2 \right. \nonumber \\
& & \left. 
~~~~ - 2 \left\{ (p^1)^2 n^2 n^3 + (p^2)^2 n^3 n^1 + (p^3)^2 n^1 n^2 \right\}
~\right]
\end{eqnarray}

Specializing to Bianchi-IX this is:
\begin{equation} \label{B-IXPot}
W_j(\vec{p})  \approx  \left\{ 
\begin{array}{ll}
- \frac{4}{ (\gamma \ell_p^2 j)^{3}} \left[~
(p^3)^2 \ (\ |p^1|^3 + |p^2|^3\ ) \ + \
(p^1)^2 \ (\ |p^2|^3 + |p^3|^3\ ) \ + \right. & \\
\left. \hspace{1.7cm} (p^2)^2 \ (\ |p^3|^3 + |p^1|^3\ ) ~\right] & 
\left(\frac{|p^I|}{\gamma \ell_p^2 j} \ll 1 \right) \nonumber \\
& \\
\frac{1}{2} \left[
\left(\frac{p^1 p^2}{p^3}\right)^2 +
\left(\frac{p^2 p^3}{p^1}\right)^2 +
\left(\frac{p^3 p^1}{p^2}\right)^2 -
2 \left\{ (p^1)^2 + (p^2)^2 + (p^3)^2 \right\} \right] & 
\left(\frac{|p^I|}{\gamma \ell_p^2 j} \gg 1 \right) \nonumber \\
\end{array}
\right.
\end{equation}

\begin{figure}
 \begin{center}
 \includegraphics[width=12cm,height=8cm]{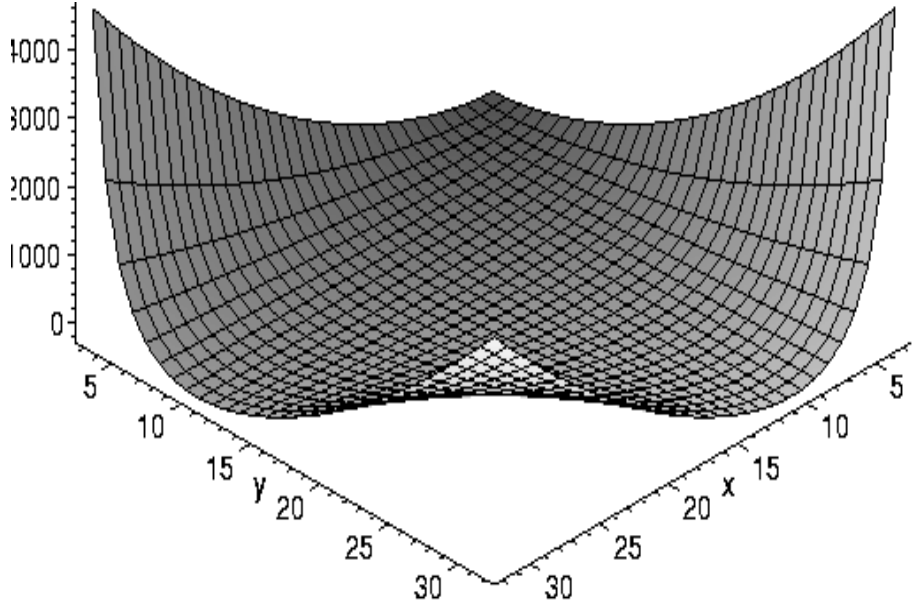}
 \end{center}
 \caption{Classical potential as a function of $x:=2p^1/\gamma\lP^2$
 and $y:=2p^2/\gamma\lP^2$ with $2p^3/\gamma\lP^2=10$.
 \label{PotClass}}
\end{figure}

\begin{figure}
 \begin{center}
 \includegraphics[width=12cm,height=8cm]{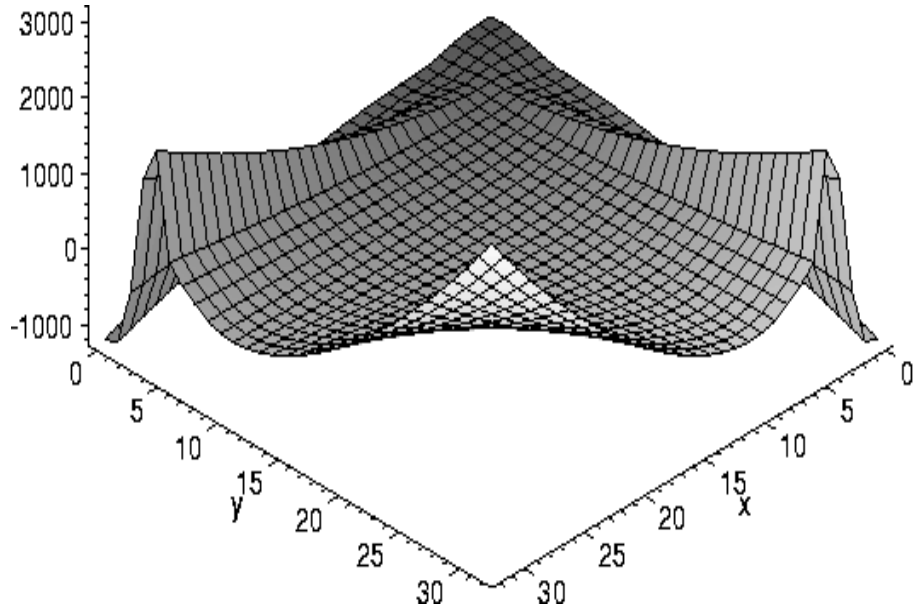}
 \end{center}
 \caption{Modified potential with collapsed walls at small $p^1$ and
 $p^2$ (with $j=3$).
 \label{Pot}}
\end{figure}

\begin{figure}
 \begin{center}
 \includegraphics[width=12cm,height=8cm]{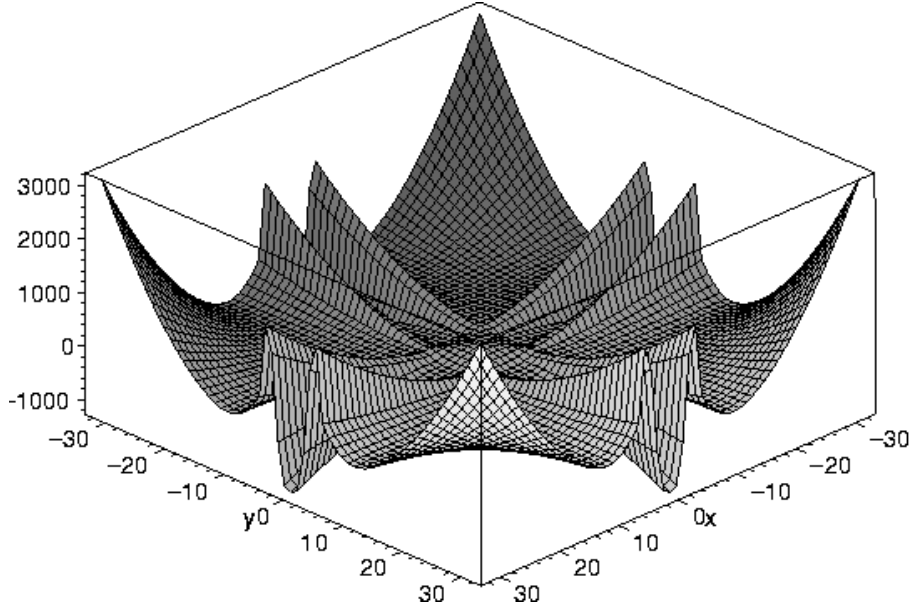}
 \end{center}
 \caption{Modified potential with collapsed walls at small $p^1$ and
 $p^2$ (diagonal canyons) with $j=3$. Also regions for negative $p^I$
 are shown to which a test particle in the potential can move; with
 the original potential only the lower quadrant is allowed.
 \label{Pot4}}
\end{figure}

A sample snapshot of the modified potential is shown in
Figs.~\ref{Pot} and \ref{Pot4} together with the classical potential
(Fig.~\ref{PotClass}) for comparison.
With the modified potential, the classical Hamiltonian
(\ref{ClassHam}) becomes
\begin{equation} \label{Heff}
\kappa N H_j^{\rm eff} (\pi_I, q^I) ~ = ~ - \frac{1}{2} ( \pi_1 \pi_2
+ \pi_2 \pi_3 
+ \pi_3 \pi_1 ) + W_j(e^{2 q^1}, e^{2 q^2}, e^{2 q^3})
\end{equation}

We have thus derived an {\em effective potential} modified by the
{\em non-perturbative} quantum effects.	Other correction
terms, such as higher order contributions to the kinetic part coming
from the discreteness, are perturbative and less important as remarked
earlier. Note that classical regime is identified by $m_I =
p^I(\frac{\gamma\lP^2}{2})^{-1} \gg 1$. The quantum modifications however 
are dominant in the {\em small $q = \frac{m}{2j}$} regimes. Choosing
larger values of the ambiguity parameter $j$, thus allows us to access the 
quantum effects even while staying in the classical regime. We can thus
study the qualitative effects of quantum modifications within a classical 
framework by using the effective potential instead of using a fully quantum 
description in terms of a wave function on minisuperspace. Effective classical
equations of motion, in terms of an external time parameter, can be derived 
from the Hamiltonian (\ref{Heff}) as we did in Section \ref{ModClass}. The 
corresponding classical trajectories can be thought of as (approximately) 
describing the motion of a quantum wave packet, giving suggestions regarding 
the full quantum behavior. 

\begin{figure}
 \begin{center}
 \includegraphics[width=5cm,height=4cm]{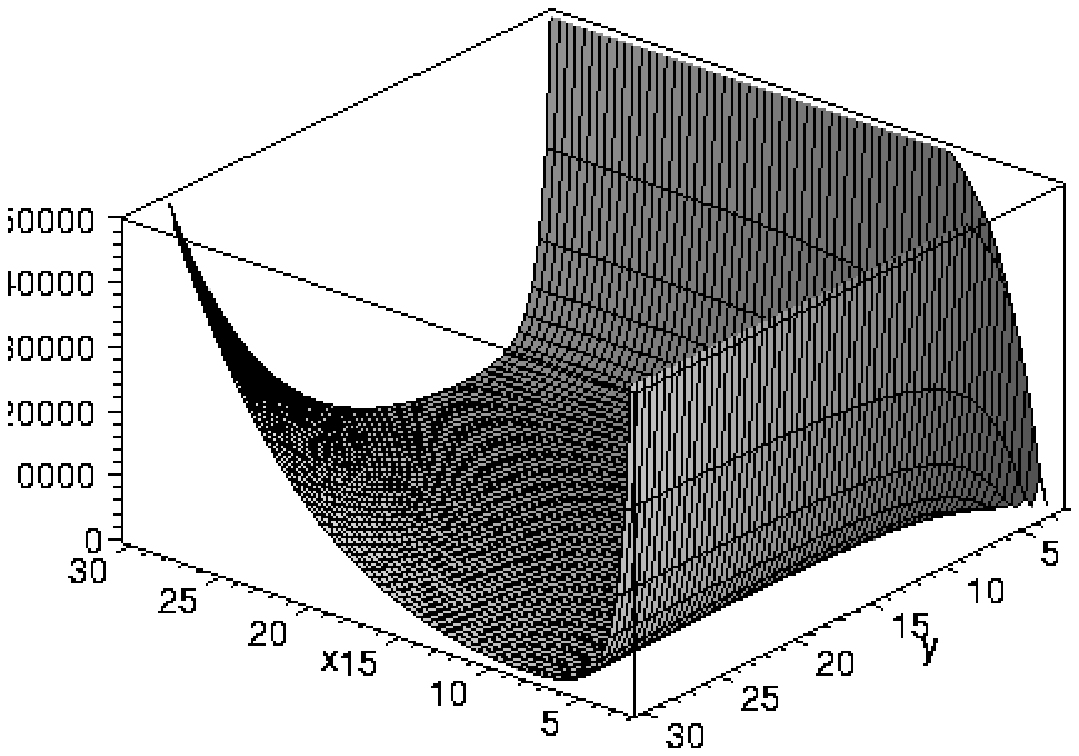}
 \includegraphics[width=5cm,height=4cm]{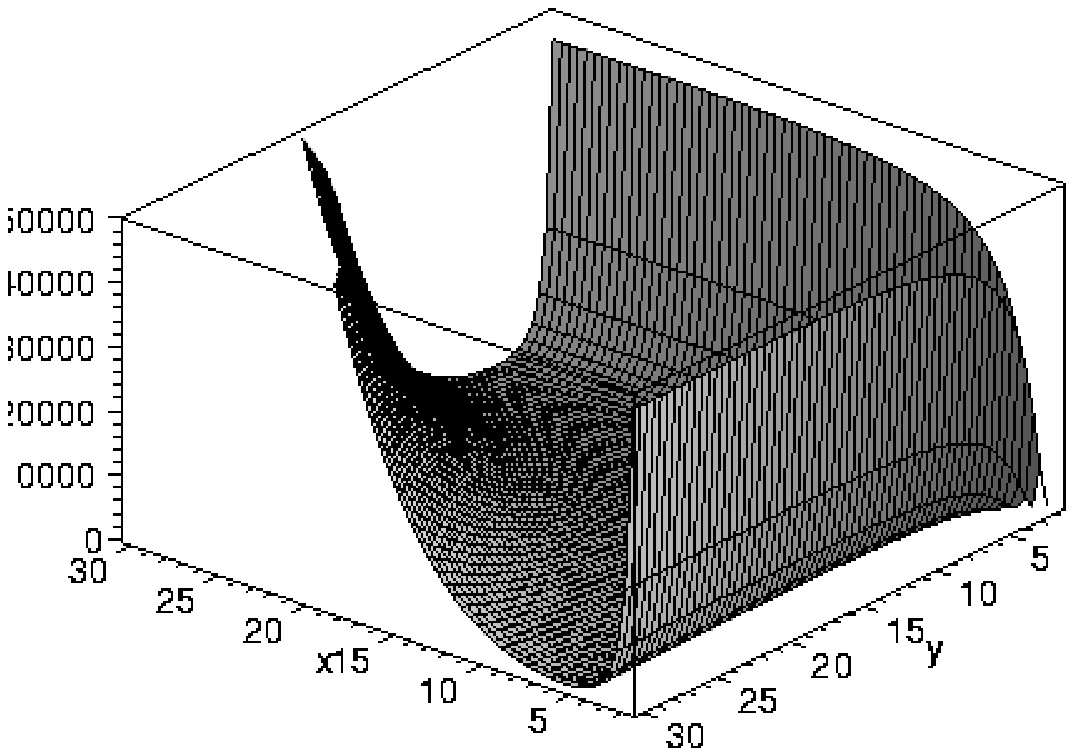}
 \includegraphics[width=5cm,height=4cm]{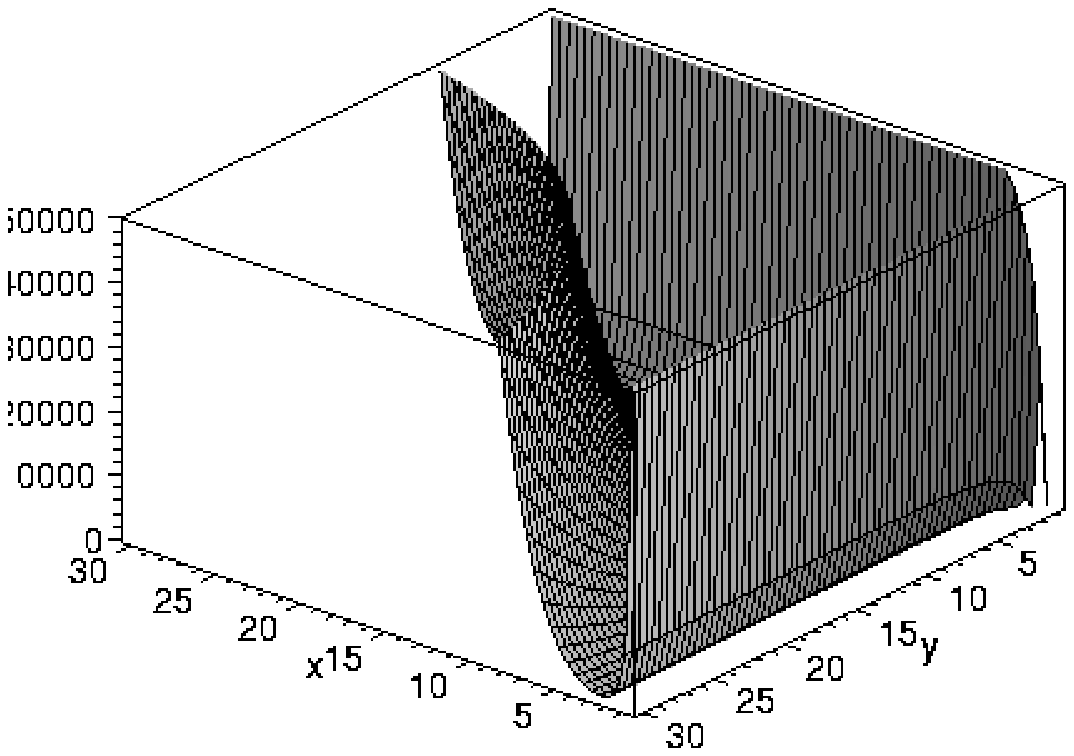}
 \end{center}
 \caption{Classical potential at constant volume
 $V_1=(\frac{13}{2}\gamma\lP^2)^{\frac{3}{2}}$,
 $V_2=(5\gamma\lP^2)^{\frac{3}{2}}$, and
 $V_3=(3\gamma\lP^2)^{\frac{3}{2}}$ (from left to right). While moving
 in the triangular valley the universe is pushed toward the classical
 singularity ($x=y=0$) by reflections on the moving left wall.
 \label{PotClassV}}
\end{figure}

\begin{figure}
 \begin{center}
 \includegraphics[width=5cm,height=4cm]{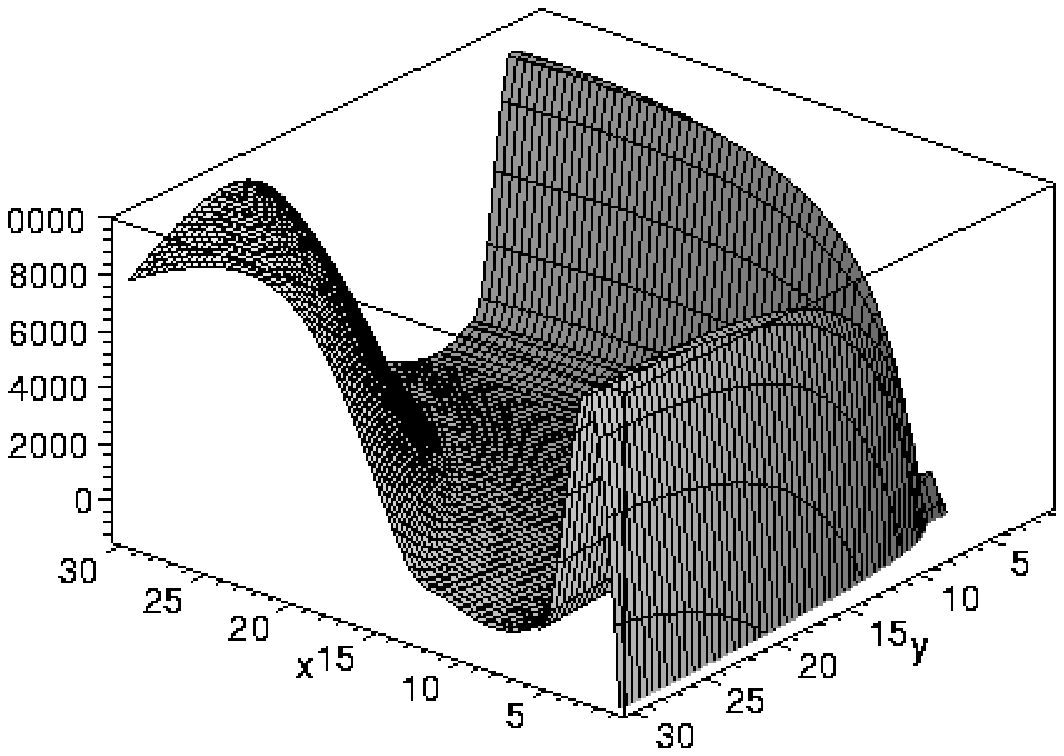}
 \includegraphics[width=5cm,height=4cm]{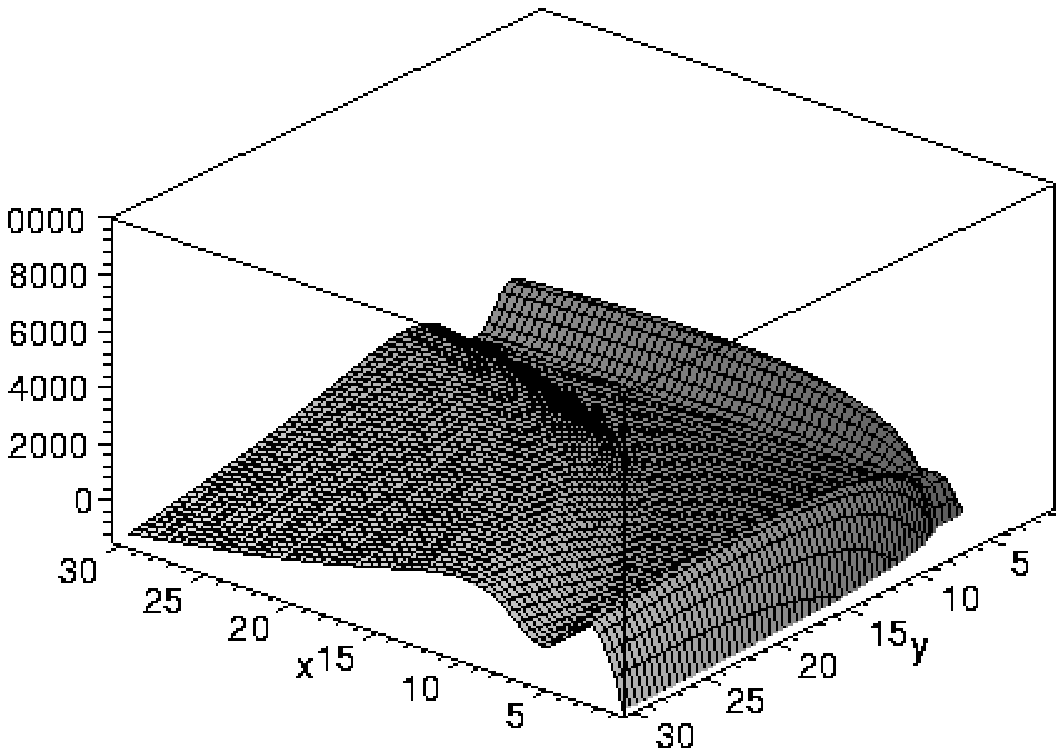}
 \includegraphics[width=5cm,height=4cm]{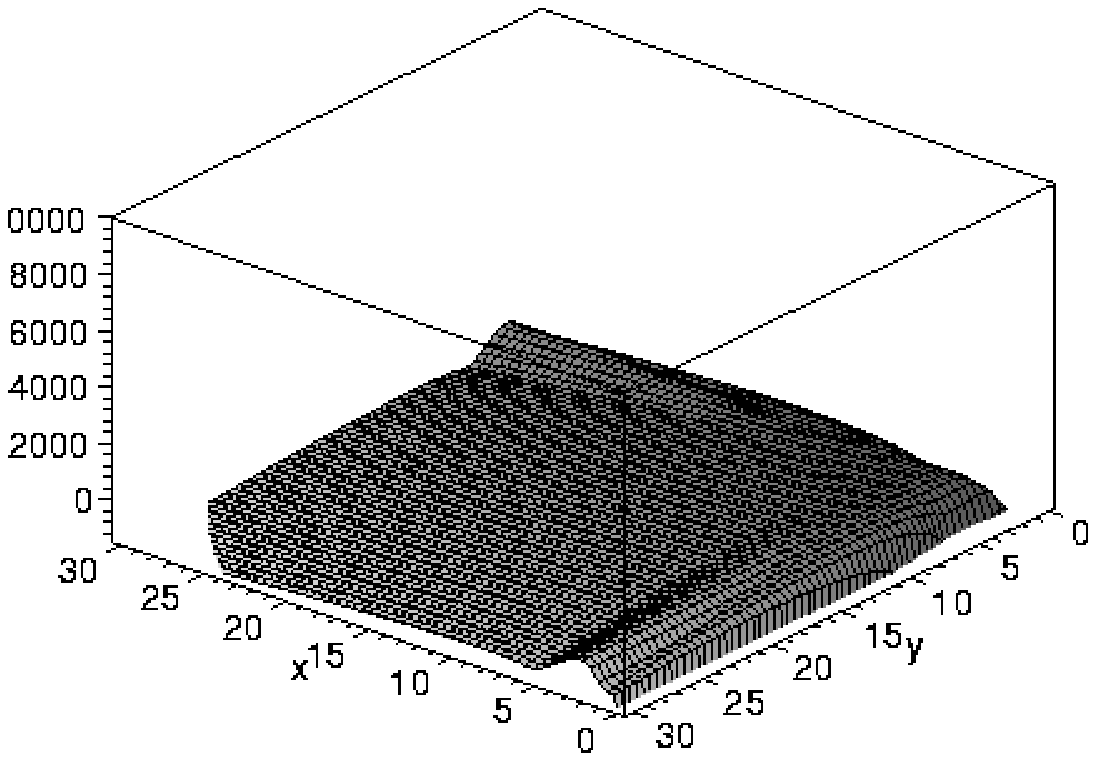}
 \end{center}
 \caption{Modified potential with collapsing walls (for $j=3$) at
 constant volume $V_1=(\frac{13}{2}\gamma\lP^2)^{\frac{3}{2}}$,
 $V_2=(5\gamma\lP^2)^{\frac{3}{2}}$, and
 $V_3=(3\gamma\lP^2)^{\frac{3}{2}}$ (from left to right).
 \label{PotVa}}
\end{figure}

Alternatively, one can also analyze the modified behavior in terms of
an internal time namely the volume. The Hamiltonian (\ref{Heff}) is in
fact hyperbolic. The potential now has a non-factorizable time dependence 
and the analysis of motion is more complicated. Figs.\ \ref{PotClassV} and 
\ref{PotVa} show the classical and the modified potential, respectively, at 
fixed volume $\sqrt{|p^1p^2p^3|}$ rather than at fixed $p^3$ as in Figs.\ 
\ref{PotClass} and \ref{Pot}. One can already expect from the figures that 
the behavior will change dramatically once the universe reaches a small triad
component. For a detailed analysis see \cite{BianchiIX,BianchiIXDetail}. 
\section{Conclusions}
\label{concl}
In this paper we have extended the methods of homogeneous loop quantum
cosmology to models with non-zero spin connection. These models have
non-trivial intrinsic curvature from the symmetric background which
requires a special treatment compared to the Bianchi I model as well as
compared to the full theory. Initially, therefore, part of the guidance from
the full theory is lost and there are more ambiguities when one
quantizes the Hamiltonian constraint. 

In \cite{Stab} two conditions which have to be imposed for a
reasonable quantization, have been introduced and discussed in
detail. These stem primarily from the requirement that the loop
quantization admit physical states which, in a semiclassical regime,
have much less sensitivity to variations on the Planck scale so that
the idealization of the continuum geometric formulation is well
justified.  This has been made more explicit by the formulation of the
continuum approximation.

In Section \ref{hami} we introduced a quantization of the Hamiltonian
constraint for all diagonalized Bianchi class A models, and showed
that the two conditions are satisfied. It was important to take the
freedom of phases in the triad representation into account which we
did by transforming to a new wave function $\tilde{s}$ in
(\ref{stilde}). The transformation was dictated by the requirement of
reproducing the Wheeler--DeWitt equation as the leading
approximation. It is satisfying to note that the same transformation
also appeared in \cite{Kodama} where it was used in the context of
relating the wave functions obtained in the Schr\"odinger quantization
based on connection variables and the wave functions of the
Wheeler--DeWitt quantization based on the metric variables. Since the
two sets of basic variables differ significantly only when the spin
connection is non-zero which happens when the intrinsic curvature is
non-zero, non-trivial phases show up in this situation. The same
transformation then appears whether one uses a loop quantization or a
Schr\"odinger quantization in the connection formulation and compares
with the Wheeler--DeWitt quantization. We also showed that the
mechanism for the absence of singularities continues to hold, despite
the phases, exactly as in the case of the Bianchi-I model
\cite{HomCosmo}. Thus, as the main result of this paper we proved
that for all diagonal Bianchi class A models there is a loop
quantization which satisfies all the conditions for a good continuum
limit as well as that of a singularity-free evolution. That such a
quantization exists at all, and even one which provides a general form
for all the models, is a non-trivial fact and gives further credence
to the viability of loop methods and the physical applications
obtained so far.

In section \ref{bianchiIX} we carried out a preliminary exploration of
the consequences of the quantized spin connection and the
correspondingly modified potential. We focused on the Bianchi IX case
and pointed out that unlike the classical potential which has infinite
walls as the singularity is approached, the quantum modified potential
has finite walls. Since the infinite walls are responsible for the
chaotic approach to the singularity, we expect that the
finite walls of the modified potential will significantly alter the
approach to the classical singularity \cite{BianchiIX,BianchiIXDetail}.

Finally a remark about the relevance of the results is in
order. Since the connection of the quantization discussed here to the
full theory is weaker, it is legitimate to question what this
quantization may have to do with the full theory. Firstly, we observe
that the methods adopted as well as the two admissibility criteria
work uniformly for all homogeneous models (including isotropic
sub-models) which do not have a non-zero spin connection. The direct
loss of connection to the full theory is only in the models with spin
connection. The treatment of the spin connection enters in two
distinct places, firstly in using the inverses of triad components and
secondly in the modification of holonomies. The first one is quite
natural and inherits the ambiguities in the definition of inverses
\cite{Ambig}. The second one has additional ambiguities due to factor
ordering. It is already clear that the first type of ambiguity is {\em
not} fixed by the two admissibility criteria. Our emphasis in this work
has been to demonstrate the existence of {\em a uniform method} of 
quantization which satisfies the two admissibility criteria {\em and} 
is singularity free. The uniqueness type of issues, while important,
requires a systematic classification of possible ambiguities and is
beyond the scope of the present work. Likewise the general question of
how many of the results of a minisuperspace quantization can survive the
full theory is also beyond the scope of this paper.

While the spin connection is seen to require a careful treatment of the
continuum approximation and also to lead to a modified approach to the
BKL singularity in the context of homogeneous models, its role in the full 
theory is more complicated and is an open issue. For instance, the phase 
factor in $\tilde{s}$ which has the form,
\[
  \exp\left(i\gamma^{-1}\lP^{-2}p^I\Gamma_I\right)=
  \exp\left(i\gamma^{-1}\lP^{-2}\smallint\md^3xE^a_i\Gamma_a^i\right) 
\]
is not even well-defined in the full theory due to the transformation
properties of the spin connection (which is no longer a covariant
object). Since the spin connection can be made arbitrarily small
locally, by choosing appropriate coordinates, one can expect that the
diffeomorphism constraint and its algebra with the Hamiltonian
constraint will play a role in this issue. Requirements such as
the local stability condition can also be expected to play a role in
analyzing this issue, at least with a local version in the absence of
a global internal time.

\begin{acknowledgments}
 M.~B.\ and G.~D.\ are grateful to Madhavan Varadarajan and Naresh
 Dadhich for invitations to and hospitality at the RRI, Bangalore, and
 IUCAA, Pune, where part of the work has been done. M.~B.\ also thanks
 the IMSC, Chennai, for hospitality. The work of M.~B.\ and K.~V.\ was
 supported in part by NSF grant PHY00-90091 and the Eberly research
 funds of Penn State.
\end{acknowledgments}

\end{document}